\documentclass[a4paper,11pt]{article}

\pdfoutput=1 

\usepackage{jheppub} 

\usepackage[T1]{fontenc}
\usepackage{amsfonts,amsmath,amssymb,epsf}
\usepackage{physics}
\usepackage{braket}
\usepackage{graphicx,color}
\usepackage{bm}
\usepackage{amstext}
\usepackage{bbm}
\usepackage{dsfont}
\usepackage{latexsym}
\usepackage{float}
\usepackage[usenames,dvipsnames]{xcolor}
\usepackage{epstopdf}
\usepackage{sidecap}
\usepackage{bbold}

\newcommand{\T}[1]{\text{#1}}

\def\la{\langle}
\def\ra{\rangle}

\newcommand{\id}{\mathds{1}}

\def\tr{{\rm Tr}}

\newcommand{\beq}{\begin{equation}}
\newcommand{\eeq}{\end{equation}}
\newcommand{\beqa}{\begin{eqnarray}}
\newcommand{\eeqa}{\end{eqnarray}}

\begin{document}

\title{Non-Hermitian Hamiltonian Deformations in Quantum Mechanics}

\author[1]{Apollonas S. Matsoukas-Roubeas,}
\author[1]{Federico Roccati,}
\author[1]{Julien Cornelius,}
\author[2]{Zhenyu Xu,}
\author[1]{Aur\'elia Chenu,}
\author[1,3]{Adolfo del Campo}

\affiliation[1]{Department  of  Physics  and  Materials  Science,  University  of  Luxembourg,  L-1511  Luxembourg,  Luxembourg}
\affiliation[2]{School of Physical Science and Technology, Soochow University, Suzhou 215006, China}
\affiliation[3]{Donostia International Physics Center,  E-20018 San Sebasti\'an, Spain}

\emailAdd{apollo.matsoukas@uni.lu}
\emailAdd{federico.roccati@uni.lu}
\emailAdd{julien.cornelius.001@student.uni.lu}
\emailAdd{zhenyuxu@suda.edu.cn}
\emailAdd{aurelia.chenu@uni.lu}
\emailAdd{adolfo.delcampo@uni.lu}

\abstract{
The construction of exactly-solvable models has recently been advanced by considering integrable $T\bar{T}$ deformations and related Hamiltonian deformations in quantum mechanics. 
We introduce a broader class of non-Hermitian Hamiltonian deformations in a nonrelativistic setting, to account for the description of a large class of open quantum systems, which includes, e.g.,  arbitrary Markovian evolutions conditioned to the absence of quantum jumps. 
We relate the time evolution operator and the time-evolving density matrix in the undeformed and deformed theories in terms of integral transforms with a specific kernel.
Non-Hermitian Hamiltonian deformations naturally arise in the description of energy diffusion that emerges in quantum systems from time-keeping errors in a real clock used to track time evolution. We show that the latter can be related to an inverse $T\bar{T}$ deformation with a purely imaginary deformation parameter. In this case, the integral transforms  take a particularly simple form when the initial state is a coherent Gibbs state or a thermofield double state, as we illustrate by characterizing the purity, R\'enyi entropies, logarithmic negativity, and the spectral form factor. 
As the dissipative evolution of a quantum system can be conveniently described in Liouville space, we further study the spectral properties of the Liouvillians, i.e., the dynamical generators associated with the deformed theories.
As an application, we discuss the interplay between decoherence and quantum chaos in non-Hermitian deformations of random matrix Hamiltonians and the Sachdev-Ye-Kitaev model.
}

\maketitle
\flushbottom

\section{Introduction}

Nonperturbative methods play a key role in physics to unveil  phenomena that do not admit an approximate  description in terms of a perturbative expansion in a small coupling constant  \cite{Marinyo15}.
A number of techniques have been developed   to describe families of  models that are solvable in a broad sense. Paradigmatic instances include Hamiltonian integrability \cite{Faddeev87}, Bethe ansatz  \cite{Takahashi99,Sutherland04,Gaudin14}, Yang-Baxter integrability \cite{Jimbo90}, quantum inverse scattering method \cite{KBI97}, quantum groups \cite{Lusztig10},  random matrices \cite{MethaBook,Forrester10}, conformal field theory \cite{DiFrancesco97}, supersymmetric methods \cite{Cooper95},  gauge-gravity dualities \cite{Ammon15}, and unitary quantum circuits, among many others. 

An important advance in this direction is the introduction of infinite families of deformations of two-dimensional integrable field theories that preserve integrability \cite{Zamolodchikov04,cavaglia2016,SmirnovZ17,jiang2021}.
In  quantum mechanics, this motivates the introduction of families of exactly-solvable Hamiltonian deformations \cite{Gross20,Gross20b,kruthoff2020,rosso21,jiang2022,ebert2022,HeXian22}.
The latter are of relevance for a system in isolation, that is  described by a time-independent Hermitian Hamiltonian with real eigenvalues.
However, the interaction of a system with the surrounding environment gives rise to decoherence \cite{Zurek03} making the system open, and no longer isolated \cite{BP02}. The evolution of the quantum state of an open system can  be described by a master equation, in which an effective non-Hermitian Hamiltonian can be identified. This prompts the consideration of non-Hermitian Hamiltonian deformations discussed in this work and the associated infinite family of solvable dissipative models.

These results should be put in a broader context aimed at finding exactly-solvable models of complex open quantum systems, which is being pursued by exploiting a variety of techniques. 
Among them, we mention exact  diagonalization of many-body Liouvillians \cite{Prosen08,Beau17},  random matrix theory \cite{Haake} including  non-Hermitian Hamiltonians and bath operators \cite{Xu19,delCampo2020,Xu21SFF,GarciaGarcia22}, random Liouvillians \cite{Can19,Sa20} and quantum channels \cite{SaProsen20}, noisy and fluctuating Hamiltonians \cite{Chenu17},  mappings between open systems and integrable systems \cite{Dukelsky21}, nonunitary quantum circuits \cite{LiChenFisher18,Skinner19,Gullans20,Ippoliti21,SaProsen21}, etc. The subclass of many-body open systems that admit a description solely in terms of non-Hermitian Hamiltonians has given rise to an emergent field, that of  non-Hermitian many-body physics, and is comparatively more developed \cite{Ashida20}. Additional efforts rely on the study of gravitational duals in the context of AdS/CFT for strongly-coupled dissipative quantum systems \cite{delCampo2020,Verlinde20, Anegawa2021,Verlinde21,Goto2021,GarciaGarcia22b}.  The use of Krylov subspace methods provides yet a different approach  \cite{KrylovDiss1,KrylovDiss2}.

From a practical point of view, Hermitian Hamiltonian deformations imply powerful identities relating the partition function and equilibrium correlation functions of the deformed and undeformed theories. In the generalized  non-Hermitian deformations we introduce here, the Hamiltonian eigenvalues are complex and their imaginary part is associated with characteristic time scales that manifest in the dynamics of the deformed theory. Non-Hermitian deformations thus imply a novel class of identities relating nonequilibrium properties of the deformed and undeformed theories. 
In particular, it is possible to relate the propagators of the deformed and undeformed theories, and thus, the corresponding time evolutions, using integral transforms with a given kernel. As specific applications, we discuss the time-dependence of the fidelity, purity, R\'enyi entropies, logarithmic negativity, and the spectral form factor (SFF) in non-Hermitian systems. For particular initial states, such as the coherent Gibbs state and the thermofield double state (TFD), these relations become particularly transparent and can be compactly expressed in terms of analytical continuations of the partition function. 
As non-Hermitian Hamiltonians can be derived from an open quantum dynamics by conditioning the evolution to the absence of quantum jumps, we apply the framework of non-Hermitian deformations to explore the role of quantum jumps in a variety of applications, including the characterization of decoherence times and the signatures of chaos in open quantum systems.

This manuscript is organized as follows. 
We first review Hermitian Hamiltonian deformations in Sec.~\ref{SecHHD} and generalize them to the non-Hermitian case in Sec.~\ref{SecNHHD}.
In Sec.~\ref{SecNHP}, we provide a short summary of the basic properties of non-Hermitian quantum dynamics.
Complex deformations are justified in the theory of open quantum systems while their relevance to physical energy-dephasing models is presented in Sec. \ref{SecED}.  
In Sec. \ref{SecSS} we study the spectral properties of the corresponding dynamical generators, presenting numerical examples from random matrix theory and the Sachdev-Ye-Kitaev (SYK) model.
In Sec. \ref{SecTFD} we study the dynamics of  a TFD, under the dynamics generated by a specific class of non-Hermitian deformations, and characterize the associated R\'enyi entropies and logarithmic negativity.
In Sec. \ref{SecSFF} we discuss the deformed SFF in the non-Hermitian setting, defined as the fidelity between an initial coherent Gibbs state and its time evolution, and use it to characterize the dynamic manifestations of quantum chaos in open quantum systems, with and without quantum jumps. 
Finally, in Sec. \ref{SecLD} we comment on possible generalizations of our results at the level of the Liouvillian and summarize our findings in Sec. \ref{secConclusions}.

\section{Hermitian Hamiltonian Deformations }\label{SecHHD}

An infinite family of exactly-solvable Hamiltonian deformations has been introduced in quantum mechanics  \cite{Gross20,Gross20b}.
In particular,  given a $d$-dimensional Hilbert space $\mathcal{H}$ and 
an isolated quantum system described by a Hamiltonian $H_0$, one considers a deformation $f(H_0)$, where $f:\mathbb{R}\rightarrow \mathbb{R}$ is a function parameterized by $\lambda \in \mathbb{R}$ with the property that
$f(H_0)\rightarrow H_0$ when $\lambda\rightarrow0$. 
In such a case,  as the original and  deformed Hamiltonians commute $[H_0,f(H_0)]=0$,
they share the same set of eigenvectors, while their eigenvalues are given by $\{E_n\}$ and $\{f(E_n)\}$, respectively.

The partition functions can be written as 
\begin{eqnarray}
Z_{0}(\beta ) &=&\int_{\mathbb{R}}\mathrm{d}E e^{-\beta E}\varrho (E), \\
Z_{f}(\beta ) &=&\int_{\mathbb{R}}\mathrm{d}E  e^{-\beta f(E)}\varrho(E)
=\int_{\mathbb{R}}\mathrm{d}E e^{-\beta E}\varrho _{f}(E), \label{eq:Zf}
\end{eqnarray}
where $\varrho(E)$ and $\varrho_f(E)$ are the density of states associated with $H_0$ and $f(H_0)$, respectively.
The two are related  by 
\begin{equation}
\varrho _{f}(E)=\varrho \big( f^{-1}(E) \big) \frac{\mathrm{d}f^{-1}(E)}{%
\mathrm{d}E},
\end{equation}%
assuming $f$ to be strictly monotonic, so that it can be inverted. 

The Boltzman factor for the deformed system can be written as an integral representation, 
\begin{eqnarray}
\label{expf}
e^{-\beta f(E)} 
&=&\int_{C_f} \mathrm{d}\beta' e^{ -\beta ^{\prime }E} K_{f}(\beta ,\beta ^{\prime })\,.
\end{eqnarray}%
When $E$ is real, it is appropriate to choose a Laplace transformation to keep the exponent real and the inverse temperature $\beta'$ on a real contour, $C_f$ being the line $\in [0, +\infty)$ or the full real axis. 
The corresponding kernel is 
\begin{equation}\label{eq:Kf}
K_{f}(\beta ,\beta ^{\prime })=\frac{1}{2\pi i}\int_{\tilde{C}_f}\mathrm{d}E \, e^{-\beta f(E)}e^{\beta ^{\prime }E}\,,
\end{equation}
with  $\tilde{C}_f$ denoting the  contour of the inverse transformation, which, for the Laplace transform, is the Bromwich contour $(\gamma-i\infty, \gamma+i\infty)$. 
It is easy to verify that the partition function of the deformed system is then related to the original, undeformed one through 
\begin{equation}\label{ZlZ0}
Z_{f}(\beta )=\int_{\mathbb{R}}\mathrm{d}E\int_{C_f} \mathrm{d}\beta' e^{
-\beta ^{\prime }E}K_{f}(\beta ,\beta ^{\prime })\varrho (E)=\int_{C_f}\mathrm{d}%
\beta ^{\prime }Z_{0}(\beta ^{\prime })K_{f}(\beta ,\beta ^{\prime })\,.
\end{equation}

In particular, we are interested in the deformation 
\begin{equation}
\label{invdef}
g (E ) = E + \lambda E^2 , \quad \lambda>0\, , 
\end{equation}%
because this spectrum originates from a non-Hermitian Hamiltonian generating  an \textit{energy dephasing} (ED) evolution \cite{Xu21SFF} in the absence of quantum jumps \cite{cornelius2022}, when the deformation parameter $\lambda$ is purely imaginary, as we discuss in Sec. \ref{SecED}.
Taking the contour $\tilde{C}_g$ as $(-\infty,+\infty)$, the kernel reads 
\begin{align}
\begin{split}
K_{g}(\beta ,\beta ^{\prime })=&\frac{1}{2\pi i}\int_{-\infty}^{+\infty}\mathrm{d}E \, e^{-\beta g(E)} e^{\beta ^{\prime }E} \\
=& \frac{1}{i \sqrt{\pi 4 \beta \lambda}} e^{\frac{(\beta-\beta^{\prime })^2}{4  \beta \lambda}}.
\end{split}
\end{align}
The deformed  partition function is then obtained from the  inverse transformation \eqref{expf} with integration on $C_g$ being the line  $(\gamma-i\infty, \gamma+i \infty)$. 
Interestingly, the inverse deformation 
\begin{equation}
    f(E) = \frac{1}{2 \lambda}\big( \sqrt{1 + 4 \lambda E}-1\big)\, , 
\end{equation} defined such that $g(f(E)) = E$, is known in the context of AdS/CFT correspondence as the 1-dimensional $T\bar{T}$ deformation \cite{Gross20,Gross20b}. Its kernel \eqref{eq:Kf} readily follows from the inverse Laplace transform 
$\mathcal{L}^{-1}[e^{-\sqrt{a s}};x] = x^{-3/2}e^{-a / (4x)} \sqrt{a }/(2 \sqrt{\pi})$ \cite{Gradshteyn2014a} that gives 
\begin{equation}
K_{f}(\beta ,\beta ^{\prime })=\frac{\beta }{\beta'\sqrt{4\pi \beta' \lambda }}e^{-\frac{(\beta -\beta^\prime)^{ 2}}{4\beta ^{\prime }\lambda
}}. 
\end{equation}
Note that the contours are related as $C_g = C_{f^{-1}}= \tilde{C}_f$ and that $C_f$ is on $[0, +\infty)$ while $\tilde{C}_{f^{-1}}$ requires the bi-lateral Laplace transform with a contour on the full real axis $(-\infty,+\infty)$.

More generally, the kernels of the original and inverse deformations are related through 
\begin{eqnarray}
    K_{f^{-1}}(\beta, \beta') &=&\frac{1}{2 i \pi} \int_{\tilde{C}_{f}}dE \, e^{-\beta f^{-1}(E)}e^{\beta' E}\nonumber \\
    &=&\frac{1}{2 i \pi} \int_{C_f}  dx \,  e^{-\beta x} f'(x) e^{\beta' f(x)}\nonumber \\
    &=&\frac{\beta}{\beta'} K_f(-\beta', -\beta),
\end{eqnarray}
as can be shown using the  change of variable $E=f(x)$  and integrating by parts, assuming that  $e^{-\beta x}e^{\beta' f(x)}$ cancels  at the edges of $C_f$. 

\section{Non-Hermitian Hamiltonian Deformations }\label{SecNHHD}
In the  context of deformations, it is also useful to relate  the propagator of the deformed Hamiltonian $U_f(t)=e^{-if(H_0)t}$ to the original one $U_0(t)=e^{-iH_0t}$. Thus, all time-evolved quantities in the deformed picture can be related to the undeformed ones. This applies to the partition function too since  $Z(\beta)=\Tr[U(-i\beta)]$.

We  assume the Hamiltonian $H_0 =\sum_n E_n \dyad{n}{\tilde n}$ is diagonal in the bi-orthogonal basis with right (left) eigenstates $\ket{n}$ ($\bra{\tilde n}$) and $E_n$ being complex in general~\cite{BrodyJPA2013}. The evolution operator reads  $U_0=e^{-i H_0 t} = \sum_n e^{-iE_nt} \dyad{n}{\tilde n}$. A generally complex deformation $w$ gives $w(H_0)=\sum_n w(E_n) \dyad{n}{\tilde n}$ and $U_w =\sum_n e^{-iw(E_n)t} \dyad{n}{\tilde n}$. In the standard case of a Hermitian Hamiltonian, right and left eigenvectors coincide and the spectrum is real.
Therefore, the relation between $e^{-iw(E)t}$ and $e^{-iEt}$ is expected to change depending on whether $E$ is real or complex.

For a Hermitian Hamiltonian $H_0$, such a relation can be derived even if the deformation is implemented by a complex function $w:\mathbb{C}\rightarrow\mathbb{C}$. For real $E$, it is appropriate to write the deformed evolution using the Fourier transform, namely 
\begin{equation}\label{simple}
	e^{-iw (E)t}
	=
	\int_\mathbb{R}  \text{d} E' \,\delta(E-E') e^{-iw (E')t}
	=
	\int_\mathbb{R}  \text{d}t'\,  K_w(t,t') e^{-iE t'}
\end{equation}
with 
\begin{equation}\label{kernel}
	K_w(t,t')
	=
	\int_\mathbb{R}  \frac{ \text{d}E}{2\pi}\, e^{i t'  E-iw (E)t}\,.
\end{equation}
We then get the relation between deformed and undeformed propagators
\begin{equation}\label{unitaryH}
    U_w(t)
    =
    \int_\mathbb{R}  \text{d}t'\,  K_w(t,t')  U_0 (t'),
\end{equation}
from which one recovers $Z_w(E)$ as in \eqref{eq:Zf} and Eq.~\eqref{ZlZ0} using the Wick rotation $i t' = \beta'$.

For a non-Hermitian Hamiltonian $H_0$ with a possibly complex spectrum, 
the relation~\eqref{simple} can be generalized as
\begin{align}\label{propNH}
	e^{-iw (E)t}
	&=
	\int_C \, \frac{\text{d} E'}{2\pi i} \frac{1}{E'-E} e^{-iw (E')t}
	=
	\int_\mathbb{R}\T{d} t'\,
	\mathcal K_w (t,t';E)
	e^{-i E t'}.
\end{align}
This kernel is obtained using $\int_0^{\infty}\text{d}t\,e^{- i s t }=1/(i s)$ for $\Re(is)>0$ to write 
\begin{align}
\begin{split}
    \frac{1}{i(E'-E)} &= \Theta(\Im(E-E'))\int_0^{\infty} \text{d}t\, e^{-i(E'-E)t} - \Theta(\Im(E'-E))\int_0^{\infty} \text{d}t\, e^{i(E'-E)t} \\
    &=\int_{-\infty}^{\infty} \text{d}t\, e^{i(E'-E)t} \big[\Theta(-\Delta_{E'})\Theta(-t) -\Theta(\Delta_{E'})\Theta(t) \big], 
\end{split}
\end{align}
that gives 
\begin{equation}
	\mathcal K_w (t,t';E)
	=
	\int_C \, \frac{\T{d} E'}{2\pi } 
	\left[
	\Theta(-\Delta_{E'}) \Theta(-t') 
	-
	 \Theta(\Delta_{E'}) \Theta(t') 
	\right]
	e^{iE' t'-i w (E')t}\,,
\end{equation}
where $\Delta_{E'} = \Im(E')-\Im(E)$ and  $C$ is an appropriate contour that includes all the eigenenergies of the original spectrum, $E$.
Notice that Eq.~\eqref{propNH} is formally equivalent to Eq.~\eqref{simple}, so that Eq.~\eqref{unitaryH} is still valid upon the replacement $K\rightarrow \mathcal K$.

Motivated by these observations, we introduce and study families of exactly-solvable and generally complex deformations of (non-)Hermitian Hamiltonians, bringing forward their use for the understanding of the effect of decoherence in chaotic quantum systems.  Eq.~\eqref{propNH} fully generalizes real deformations of Hermitian Hamiltonians to complex ones and to non-Hermitian Hamiltonians. However, the applications we will consider in Sec.~\ref{appli} are based on the energy dephasing channel which can be understood as a non-Hermitian deformation of a Hermitian Hamiltonian. Therefore, Eq.~\eqref{simple} will be the most relevant in the following.

\section{Non-Hermitian Dynamics}\label{SecNHP}

One of the fundamental postulates of quantum physics states that an isolated system is described by a Hermitian Hamiltonian. As a result, the dynamics it generates is described by a unitary  time-evolution operator.
Given the state of an isolated system, this assures the conservation of probability in the measurement outcomes and restricts the expectation value of energy to the real numbers.
Nevertheless, since the very early days of quantum theory \cite{gamow1928,majorana1928}, numerous heuristic attempts to account for dissipative phenomena in nuclear, atomic and molecular physics,  have employed effective non-Hermitian Hamiltonians \cite{Ashida20,moiseyev2011}. 
In the last two decades, the proposal of parity-time $(\mathcal{PT})$ symmetry as an alternative to Hermiticity \cite{BenderPRL1998,bender2007} paved the way for the systematic study of non-Hermitian physics. 
By now, it is understood that non-Hermitian Hamiltonians can be rigorously justified when the dynamics is restricted to a subspace of interest (e.g., making use of projection operator methods) and in the context of quantum measurement theory, by conditioning quantum trajectories on given measurement outcomes \cite{Ashida20}. 

Starting from the
Schr\"odinger equation with a  non-Hermitian Hamiltonian $H$ 
\begin{equation}\label{schr}
    i \partial_t \ket{\psi (t)} = H \ket{\psi (t)}
\end{equation}
one gets
\beqa \label{NHschr1}
\partial_t \rho (t) = -i\left(H \rho (t) - \rho (t) H^\dag \right)  
\eeqa
for the corresponding density matrix $\rho (t) = \ketbra{\psi(t)}$.

One can always decompose the Hamiltonian $H$ into a sum of a Hermitian and an anti-Hermitian term
\beqa \
H= H_0 - i \Gamma_0 ,
\eeqa
where $H_0=\frac{1}{2}(H+H^\dag)$ and $\Gamma_0 =\frac{1}{2i}(H^\dag -H)$ are Hermitian.
Then, the non-Hermitian evolution   \eqref{NHschr1} becomes
\beqa \label{jfGKLS0}
\partial_t \rho (t)=-i[H_0,\rho (t)] - \{ \Gamma_0, \rho (t) \} ,
\eeqa
involving only a commutator for the Hermitian part $H_0$ and an anti-commutator arising from the anti-Hermitian part $-i\Gamma_0$.
We note that under such dynamics, the trace of the density matrix is in general not preserved.
Nevertheless, one can enforce the property $\tr[ \rho (t) ] = 1 $, starting from a normalized initial state, by the addition of a term involving a time-dependent coefficient
\begin{equation}
\label{chit}
    \chi (t) = 2  \tr[ \Gamma_0  \rho (t)]\,.
\end{equation}
In such a scenario, the dynamics is generated by the nonlinear 
equation~\cite{BrodyGraefe12}
\beqa
\partial_t \rho (t) = -i\left[H_0, \rho (t) \right] - \{ \Gamma_0 ,  \rho (t) \} + 2 \tr[ \Gamma_0 \rho (t) ]   \rho (t) ,
\label{rhotDBNGL}
\eeqa
with general analytic solution 
\beqa
\rho (t) = \frac{  e^{-i  Ht}  \rho (0)  e^{i  H^\dagger t} }{ \tr \left[e^{-i  Ht}  \rho (0)  e^{i  H^\dagger t}   \right]  }  \,.
\label{solutionBNGL0}
\eeqa

This kind of evolution characterized by balanced norm gain and loss (BNGL) is known to arise in $\mathcal{PT}$-symmetric quantum mechanics \cite{BrodyGraefe12}. In the context of continuous quantum measurements, the above equation is also known as the nonlinear Schr\"odinger equation for null-measurement conditioning 
\cite{Carmichael09}. In addition,  it has recently been pointed out that an {\it arbitrary} evolution characterized by a time-dependent density matrix $\rho(t)$ admits an equation of motion characterized by BNGL dynamics \cite{Alipour2020}.

We observe here that  pure states remain pure under BNGL dynamics. Indeed,
an initial pure state $\rho(0) = \ketbra{\psi} $,  under the evolution  \eqref{rhotDBNGL} has R\'enyi entropy
\begin{equation}
S_\alpha [\rho(t)] =  
 \ln \big( \tr [  \rho(t)^\alpha ]  \big) /( 1- \alpha) = 0,    
\end{equation}
with $\alpha>0,  \alpha \neq 1 $. 

For the undeformed Hermitian Hamiltonian $H_0$ the evolution given by BNGL dynamics simply yields $\tilde\rho_0 (t)=U_0(t)\rho(0)U_0^\dagger(t)$, as the trace is preserved. 
As shown in Appendix \ref{appFid}, the corresponding BNGL dynamics generated by the deformed Hamiltonian $w(H_0)$ is given in terms of $\tilde\rho_0 (t)$ as  
\begin{equation}\label{evolstate}
    \rho_w(t)
    =
    \frac{\int_\mathbb{R}  \text{d}s \, \int_\mathbb{R}  \text{d}s'\,   K_w(t,s) K_{w^*}(-t,-s')  U_0 (s-s')\tilde \rho_0(s')}{\int_\mathbb{R}  \text{d}s \, \int_\mathbb{R}  \text{d}s'\,   K_w(t,s) K_{w^*}(-t,-s')\Tr[  U_0 (s-s')\tilde \rho(s')]}\,.
\end{equation}

In order to motivate the BNGL equation, which we will use throughout the manuscript, we shortly describe here its connection with the standard Lindblad dynamics for open quantum systems.

The embedding of a quantum system in a surrounding environment makes its dynamics open and not unitary.
The time evolution of an open quantum system is generally described by a master equation of the form
\cite{BP02}
\beqa \label{MasterEqGeneral}
\partial_t \rho (t)=-i[H_0,\rho (t)]+\mathcal{D}[\rho (t)],
\eeqa
where  $H_0$ is the system Hamiltonian (including the Lamb shift) and 
the breaking of unitarity is induced by the dissipator $\mathcal{D}[\cdot]$, which accounts for the interaction with the environment. 
A seminal result in the theory of Markovian open quantum systems is that the evolution is described by the Gorini–Kossakowski–Sudarshan–Lindblad (GKLS) equation that admits the canonical Lindblad form  \cite{GKS76,Lindblad76,BP02}
\beqa \label{GKLSgeneral}
\partial_t \rho (t)=-i[H_0,\rho (t)]
+
\sum_\alpha 
\gamma_{\alpha}\left(L_\alpha \rho (t) L_\alpha^\dag - \frac{1}{2} \{ L_\alpha^\dag L_\alpha, \rho (t) \}\right),
\eeqa
where $\gamma_\alpha\geq 0$ are the time-independent coefficients, 
$L_\alpha$ are the jump operators,  and $H_0$ is the Hermitian system Hamiltonian. 
In the quantum jump approach \cite{Carmichael09}, it is customary to rewrite the above equation as
\beqa
\partial_t \rho (t) =-i\left(H_{\rm eff} \rho (t) - \rho (t) H_{\rm eff}^\dag \right)+
\sum_\alpha 
J_\alpha \left[\rho (t)\right],
\eeqa
in terms of the effective non-Hermitian Hamiltonian
\begin{equation}   \label{Heff}
H_{\mathrm{eff}}=H_{0}-\frac{i}{2}
\sum_\alpha 
\gamma _{\alpha }L_{\alpha
}^{\dag }L_{\alpha }  ,
\end{equation}%
and the jump superoperators
\beqa
J_\alpha \left[\rho (t)\right]
=
\gamma_\alpha L_\alpha \rho (t) L_\alpha^\dag .
\eeqa
In the context of continuous quantum measurements 
the Lindblad master equation can be seen as the unconditional dynamics of the system, that is as an average over all possible trajectories in which quantum jumps take place. 
One can interpret then the BNGL evolution 
conditioned on the absence of  quantum jump, i.e., disregarding the contribution from $J_\alpha \left[\rho (t)\right]$ for a subensemble of trajectories~\cite{Carmichael09,MingantiPRA2019,RoccatiOSID2022}.
The time-evolution is then exclusively governed by the non-Hermitian Hamiltonian
\beqa  \label{NHschr0}
\partial_t \rho (t) =-i\left(H_{\rm eff}  \rho (t) - \rho (t) H_{\rm eff}^\dag \right).
\eeqa
Upon normalization, the dynamics becomes trace-preserving and  leads to the BNGL equation (\ref{rhotDBNGL}) with $\Gamma_0=\frac{1}{2}
\sum_\alpha 
\gamma _{\alpha }L_{\alpha
}^{\dag }L_{\alpha }$.

\section{Applications of Non-Hermitian Deformations}\label{appli}

In this section we will introduce and study in detail the energy dephasing channel. This can be described as a complex, i.e.,~non-Hermitian, deformation of an Hermitian Hamiltonian.

\subsection{Energy Dephasing and Decoherence Time}
\label{SecED}

We consider here the simplest energy dephasing (ED) model, i.e., a dissipator with a single jump operator $L_\alpha=L_\alpha^\dag=H_0$. 
In this case, the Lindblad form (\ref{GKLSgeneral}) reduces to the master equation describing energy diffusion
\beqa
\partial_t\rho&=&-i[H_0,\rho]-\gamma[H_0,[H_0,\rho]]\,, \quad \gamma>0.
\label{rhotED}
\eeqa
%
We note that the quantum state under energy-dephasing   (\ref{rhotED})  evolves as
\beqa \label{rhotED2}
\rho(t)=\sum_{nm}\rho_{nm}(0)e^{-i(E_n-E_m)t-\gamma t (E_n-E_m)^2}|n\ra \la m|.
\eeqa
To characterize the role of decoherence during the time evolution, we consider the purity $P(t)=\tr[\rho(t)^2]$, which is related to the R\'enyi-2 entropy $S_2[\rho(t)]$ as $P(t)=e^{-S_2[\rho(t)]}$. 
In the case of ED, it reads
\beqa
\label{PtED}
P(t)=\sum_{nm}\rho_{nm}(0)^2e^{-2\gamma t (E_n-E_m)^2}.
\eeqa

The corresponding evolution in the absence of quantum jumps is given by the BNGL equation (\ref{rhotDBNGL}) with the deformed Hamiltonian
\begin{equation}\label{deform2}
    w(H_0)=H_0-i\gamma H_0^2\,,
\end{equation}
for which the kernel in Eq.~\eqref{kernel} reads
\begin{equation}\label{dephkernel}
    K_w(t,s) = \sqrt{\frac{\pi}{\gamma t}} e^{-(t-s)^2/4\gamma t}\,.
\end{equation} 
With this kernel, the propagator of the deformed theory, and therefore the time evolution under BNGL, can be explicitly found.
We note that the deformation~\eqref{deform2} is equivalent to the inverse  $T\bar{T}$-deformation
in Eq. (\ref{invdef}) with a purely imaginary value of the deformation parameter $\lambda=-i\gamma$.
Making use of Eq.~\eqref{evolstate},
the explicit expression of the evolved state under BNGL energy dephasing  is
\beqa
\label{rhotBNGL2a}
\rho(t)=\frac{\sum_{nm}\rho_{nm}(0)e^{-i(E_n-E_m)t-\gamma t (E_n^2+E_m^2)} |n\ra \la m| }{\sum_{j}\rho_{jj}(0)e^{-2t\gamma E_j^2}},
\eeqa
and the corresponding time-dependent  purity equals
\beqa
\label{PtEDBNGL}
P(t)=\tr[\rho(t)^2]=\frac{\sum_{nm} \abs{\rho_{nm}(0)}^2e^{-2\gamma t (E_n^2+E_m^2)}}{(\sum_{n}\rho_{nn}(0)e^{-2t\gamma E_n^2})^2}.
\eeqa
The last expression has the remarkable feature that whenever the initial state is pure, such that the  factorization $\rho_{nm}(0)=c_n(0)c_m(0)^*$ holds, then $P(t)=1$ (equivalently as shown in Sec. \ref{SecNHP}, $S_2[\rho(t)]=0$). Thus, the evolution preserves the purity of a pure quantum state, even when it exhibits a dissipative evolution. In particular, in this case, Eq. (\ref{rhotBNGL2a}) reduces to
\beqa
\label{rhotBNGL2a1}
\rho(t)=\frac{\sum_{nm}c_{n}(0)c_{m}(0)^*e^{-i(E_n-E_m)t-\gamma t (E_n^2+E_m^2)}}{\sum_{n}|c_n(0)|^2e^{-2t\gamma E_n^2}}|n\ra \la m|.
\eeqa
Comparison of the time evolution under ED (\ref{rhotED2}) and BNGL (\ref{rhotBNGL2a})  reveals the role of quantum jumps. The latter becomes particularly transparent by analyzing the decoherence time, that can be derived from the purity, as we next show.
Specifically, for an initial mixed state, the  decoherence time $\tau_D$  can be extracted from the short-time decay of the purity \cite{Beau17} 
\beqa
P(t)=P(0)\left[1-\frac{t}{\tau_D}\right]+\mathcal{O}(t^2).
\eeqa
For an arbitrary Markovian evolution described by a Lindblad master equation, the decoherence time is given by the inverse of the covariance of the Lindblad operators evaluated in the initial state \cite{Chenu17}. 
 For an initial mixed state evolving under ED, it is set by the inverse of the energy fluctuations in the initial state \cite{Xu19,delCampo2020,Xu21SFF}
\beqa
\frac{1}{\tau_D}=\frac{4\gamma}{P(0)} \left(\tr[\rho(0)^2H_0^2]-\tr[\rho(0)H_0\rho(0)H_0]\right).
\eeqa
We note that $1/\tau_D\geq0$ and it vanishes only when the initial state is diagonal in the Hamiltonian eigenbasis, $[\rho(0), H_0]=0$. This latter case includes the possibility that the initial state is a (pure) eigenstate of $H_0$ or that $\rho_0$ is a mixed equilibrium state.
By contrast, in the case of BNGL, the decoherence time reads
\beqa
\frac{1}{\tau_D}=\frac{4\gamma}{P(0)} \left(\tr[\rho(0)^2H_0^2]-P_0\tr[\rho(0)H_0^2]\right).
\eeqa
 This expression identically vanishes when the initial state is a pure state describing an arbitrary coherent superposition of energy eigenstates, i.e., when $P(0)=1$ and $\rho(0)^2=\rho(0)$. In addition, an initial mixed state that is diagonal in the energy eigenbasis  has finite $\tau_D$ and evolves nontrivially under BNGL. In short, the absence of quantum jumps associated with non-Hermitian deformations alters the value of the decoherence rate and changes the conditions under which it vanishes.

\subsection{Spectral Structure of the Dynamical Generators }\label{SecSS}

Quantum chaos has historically been founded upon the study of complex Hamiltonian spectra, which in principle contain all the information required to describe the evolution of an isolated system \cite{berry1977, bohigas1984, Haake}. 
At the same time, a robust theory of open quantum systems has been established, focusing on the overall dynamical maps that control the temporal evolution of a subsystem \cite{BP02}. 
Thus, the study of the spectral properties of complex non-Hermitian dynamical generators and maps is of great relevance to the investigation of the fate of the signatures of quantum chaos in open dynamics \cite{feinberg1997,wang2020,Sa20,marinello2016,mochizuki2020, wang2020a,Can19,can2019a,denisov2019}.

In order to study the spectral properties of the dynamical generators discussed in the previous sections,  we fix a vectorization process for all elements of the space of  the density matrices,  $ \mathcal{H}^* \otimes \mathcal{H}$.   
The Hilbert space of all linear superoperators acting on the density matrices is often referred to as Liouville space.  
The Liouville space formalism is extensively used for the study of the spectral properties of quantum channels and open systems \cite{Haake,Gyamfi20}.  
The properties of vectorized matrices have to be  treated carefully, as the vectorization process is basis dependent.  

Let  $\big\{ \ket{i}  | \,  i \in  \{ 1,2, \dots ,  d \} \big\}$ be the complete eigenbasis of the undeformed Hamiltonian $H_0$ for the Hilbert space $\mathcal{H}$. 
Any linear operator $ \rho \dot{=} \{ \rho_{ij} \}$ can be represented as a vector 
	\begin{equation}
	\label{vectorization}
\mathcal{\rho} = \sum_{i,j=0}^{d-1} \rho_{ij} \ketbra{i}{j}  \rightarrow   | \rho ) = \sum_{i,j=0}^{d-1} \rho_{ij} \ket{i} \otimes \ket{j}^* .
	\end{equation}
For this specific choice of horizontal vectorization,  any set of linear operators $A \dot{=} \{ a_{ij} \}$,  $B \dot{=} \{ b_{ij} \}$ acting on the vectorized operator $\rho$ from left and right respectively,  can be represented as a superoperator with the use of the Kronecker product $\otimes$ of $A$ and the transpose $B^\intercal$,
\begin{equation}
A \rho B \rightarrow ( A \otimes B^\intercal )  | \rho ) .
\end{equation}
For example, the Liouvillian which generates the unitary evolution of the Hermitian Hamiltonian $H_0$ is represented as
\beqa \label{isoliouv}
\mathcal{L} = - i [ H_0,  \cdot ]
\; \rightarrow \;
\mathbb{L} = -i ( H_0 \otimes \id - \id \otimes H_0^\intercal )   .
\eeqa

In what follows,  we shall study the spectral properties of the generators of the non-Hermitian deformations discussed in the previous sections,  
relating them to the ones of the associated energy dephasing double bracket Lindbladians. 
Specifically, we will see that ED models have their spectrum on a one dimensional locus, reflecting the freedom in the choice of the ground state of $H_0$.
The removal of the quantum jump term, which leads to the associated non-Hermitian deformation BNGL model, spreads the spectrum in an area determined by the deforming function. The eigenvalue density on the complex plane is then rigidly shifted with time by $\chi(t)$ in Eq. (\ref{chit}).

For the sake of illustration, 
we consider as well the more general complex deformation  
\begin{equation}\label{deformk}
    w(z)=z-i\gamma z^\kappa\,.
\end{equation} 
We highlight here some properties that will be useful in the following.

The state at time $t$ is given by
\begin{equation}
\rho (t)=\ \sum_{n,m=1}^{d}\frac{\rho
_{nm}(0) e^{-i (E_{n}-E_{m}) t- \gamma \left( E_{n}^{\kappa
}+E_{m}^{\kappa }\right) t}}{\sum_{j=1}^{d}\rho _{jj}(0)e^{-2\gamma
E_{j}^{\kappa }}}
\dyad{n}{m}\,.
\end{equation}
The corresponding Liouvillians in the vectorized formalism read
\begin{equation}
\mathbb{L}^{(\kappa )}=i(\id \otimes H_{0}^\intercal - H_{0}\otimes \id)-\gamma \left( \id\otimes (H_{0}^\intercal)^{\kappa }+H_{0}^{\kappa
}\otimes \id \right) + 2 \gamma \mathrm{Tr}(H_{0}^{\kappa }\rho
(t))\left( \id\otimes \id\right) ,   \label{liouvilleBNGLk}
\end{equation}%
and satisfy the eigenvalue equation
\begin{equation}
\mathbb{L}^{(\kappa )}  | n,  m )
= \left( i(E_{n}-E_{m})-\frac{\gamma }{2}(E_{n}^{\kappa }+E_{m}^{\kappa
})+ 2 \gamma \mathrm{Tr}(H_{0}^{\kappa }\rho (t))\right)  | n,  m ) , \label{eigBNGLk}
\end{equation}
denoting $| n,  m ) \equiv \ket{n} \otimes \ket{m}^*$.
For Hamiltonians with a bounded spectrum (that is, included in the interval $E\in (-R,R)$, with $R>0$), the spectrum of $\mathbb{L}^{(\kappa )}$ is
bounded, for even $\kappa$, by the boundaries of the three functions
\begin{equation}
\lambda _{b}=
\begin{cases}
-\gamma E^{\kappa }  -2 i E + 2 \gamma \mathrm{Tr}(H_{0}^{\kappa }\rho (t)) \\
-\frac{\gamma }{2}(R^{\kappa } +E^{\kappa })+i(R-E) + 2 \gamma \mathrm{Tr}(H_{0}^{\kappa }\rho (t))   , \\
-\frac{\gamma }{2}(R^{\kappa }+E^{\kappa })-i(R+E) + 2 \gamma \mathrm{Tr}(H_{0}^{\kappa }\rho (t)) 
\end{cases}
\label{boundeven}
\end{equation}
and  by the boundaries of the four functions
\begin{equation}
\lambda _{b}=\pm \frac{\gamma }{2}(R^{\kappa }+E^{\kappa })\pm i(R-E) + 2 \gamma \mathrm{Tr}(H_{0}^{\kappa }\rho (t)) ,    \label{boundodd}
\end{equation}
for odd $\kappa$.


When $\kappa$ is even, one can construct the vectorized canonical Lindblad forms of Eq. \eqref{GKLSgeneral}, having the $k$th power of the undeformed Hamiltonian as a single Lindblad operator
\begin{align}
\mathbb{L}_{\text{ED}}^{(\kappa)}=i(\id\otimes H_{0}^\intercal - H_{0}\otimes \id%
)-\gamma \left( \id\otimes (H_{0}^\intercal)^{\kappa}+H_{0}^{\kappa} \otimes \id
-2 H_{0}^{\frac{\kappa}{2}} \otimes (H_{0}^\intercal)^{\frac{\kappa}{2}} \right) ,  \label{liouvilleED}
\end{align}
and compare their spectra with the corresponding generators, after removing the quantum jumps in Eq. \eqref{liouvilleBNGLk}. 

Considering the case of BNGL when $\kappa=2$,
we see that the spectrum of the ED Liouvillian $\mathbb{L}_{\text{ED}}^{(2)}$ only depends on the energy gaps $%
E_{nm}\equiv E_{n}-E_{m}$, laying on the parabola $\lambda =-\frac{\gamma }{2%
}E_{nm}^{2}+iE_{nm}$ with a probability distribution given from the density
of gaps of the Hamiltonian $H_{0}$. 
Neglecting the time-dependent shift of the spectrum, when the jump term is removed, the eigenvalues $\lambda $ are spread on a two
dimensional locus defined by the boundaries of the three  parabolas
\begin{equation}
\lambda _{b}=%
\begin{cases}
-\gamma E^{2} -2 i E \\
-\frac{\gamma }{2}(R^{2}+E^{2})+i(R-E) \\
-\frac{\gamma }{2}(R^{2}+E^{2})-i(R+E)%
\end{cases}%
,\quad E\in (-R,R)\,,  \label{bound}
\end{equation}%
where $R$ is the largest allowed eigenvalue. 
For  simplicity, one can always consider  the spectrum of $H_{0}$ to be distributed in the interval $(-R,R)$.  
Every eigenvalue of the $\mathbb{L}^{(\kappa)}$ spectrum can be thought of as a point on a shifted parabola of the corresponding ED spectrum,  centered on itself,  within  the domain $(-R,R)$.  
Finally, the inclusion of a time-dependent and initial condition-dependent term in the Liouvillian $\mathbb{L}^{(2)}$ of Eq. \eqref{liouvilleBNGLk}
shifts the spectrum on the real axis by $\chi (t) = 2 \gamma \tr[ H_{0}^{2}\rho (t) ] $.

\paragraph{Examples from Random Matrix Theory.}

\begin{figure}
\begin{center}
\hspace{-0.5cm}
\includegraphics[width=1 \linewidth]{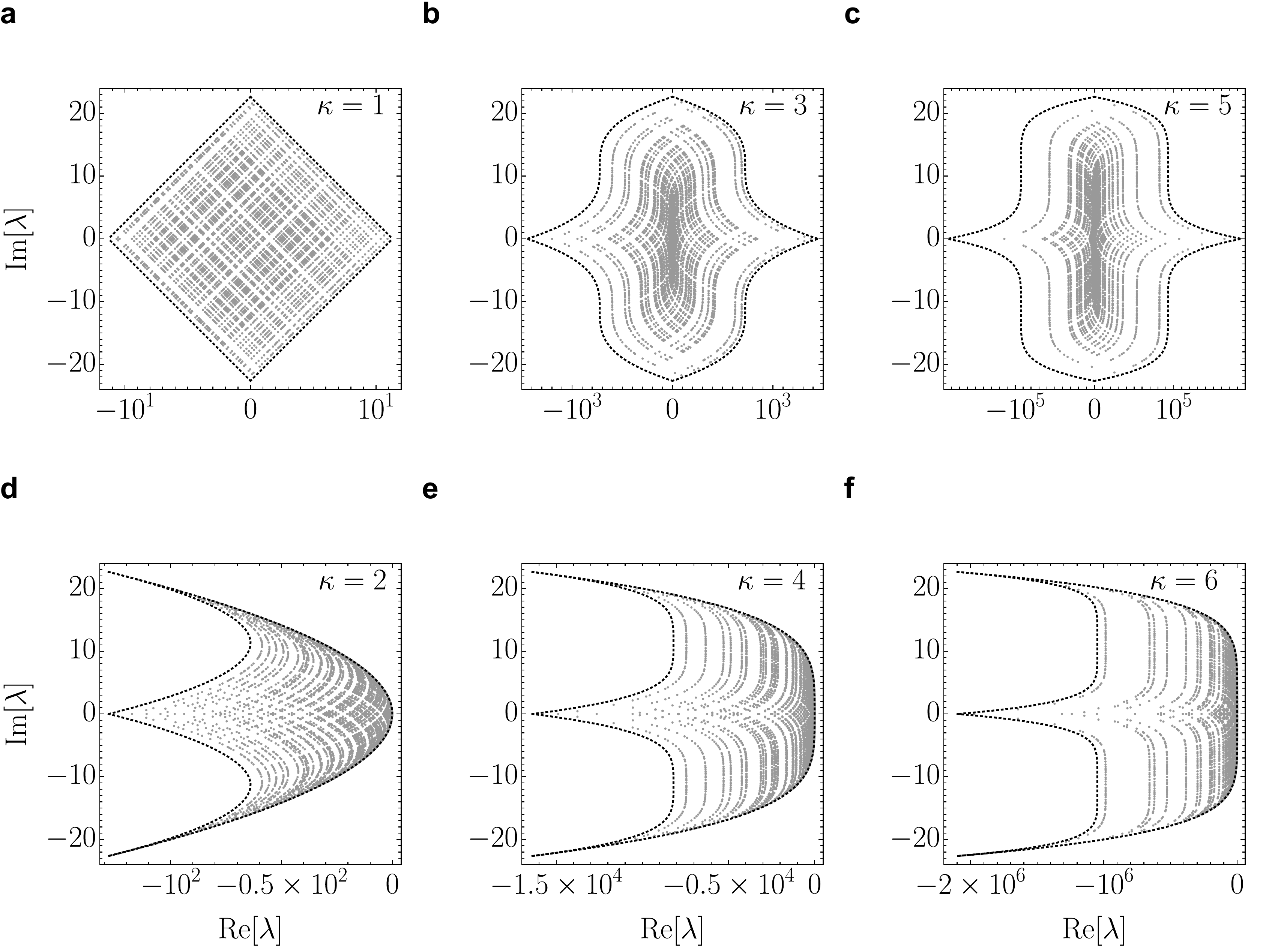}
\end{center}
\vspace{-0.5cm}
\caption{\label{Fig1DecoTTbar} {\bf  Liouvillian spectra of different non-Hermitian deformations. }
 Spectra of $\mathbb{L}^{(\kappa )}$ with $\gamma=1$ when $%
t\rightarrow \infty $, for $\protect\kappa =1,2,3,4,5,6$ (gray points) on
the complex plain, together with the theoretical boundaries (dashed black lines)
given by Eq. \eqref{boundeven} and Eq. \eqref{boundodd}.   In all six plots we show the spectrum of a single random $\mathrm{GOE}(2^6)$ Hamiltonian $H_{0}$ with $%
\protect\sigma =1$.
}
\end{figure}

Since Wigner's groundbreaking work on the neutron excitation spectra of heavy nuclei \cite{wigner1955}, it has become clear that random matrices can adequately describe the statistical features of several quantum systems \cite{MethaBook,Haake}.
As paradigms of quantum chaotic Hamiltonians with bounded spectrum and time-reversal symmetry, we sample random $d$-dimensional matrices from the Gaussian orthogonal ensemble,  $H_0\in{\rm GOE}(d)$.  
Specifically, we consider samples of real,  orthogonal matrices $H=(X+X^{\intercal })/2$,
where all elements $x\in \mathbb{R}$ of $X$ are pseudo-randomly generated
with probability measure given by the Gaussian $\frac{1}{\sigma \sqrt{2\pi }} e^{-\frac{x^{2}}{2\sigma ^{2}}}$  with standard deviation $\sigma$ \cite{MethaBook,Haake}. 
When $d$ or the sample size is large, the spectral density distribution of such matrices can be approximated by the semicircle law
\begin{equation}
c(E)=\frac{\sqrt{2d\sigma ^{2}-E^{2}}}{\pi d\sigma ^{2}}
\label{semicircle}
\end{equation}
so $R=\sigma \sqrt{2d}$.
In Fig.~\ref{Fig1DecoTTbar} we show the BNGL spectra of $\mathbb{L}^{(\kappa )}$ when $t\rightarrow \infty $ for $\kappa =1,2,3,4,5,6$ with the corresponding theoretical boundaries of Eq.  \eqref{boundeven} and \eqref{boundodd}. 
Each panel has the spectrum of a Hamiltonian drawn from $\mathrm{GOE}(64)$, with $\sigma=1$.

%

\paragraph{Examples from the Sachdev-Ye-Kitaev Model.}

For the illustration of the relation between BNGL and ED Liouvillian spectra,  we consider the example of a Hilbert space of dimension $d = 2^ N$ and the undeformed SYK Hamiltonian of $2 N$ Majorana Fermions with an all-to-all random quartic interactions in the occupation number representation
\begin{align}
\label{hamilton}
 H_0 = \frac{1}{4!} \sum_{k,l,m,n=1}^{2N}  J_{klmn} \chi_k  \chi_l  \chi_m  \chi_n  ,
\end{align}
obeying the anti-commutation relation $ \{\chi _{k},\chi _{l}\}=2\delta _{kl}$.
The factor of two in the latter can be seen as a rescaling of
the operators \cite{Solano2017,Garcia-Garcia2017PRD}. 
The coupling tensor $J_{klmn}$ is completely anti-symmetric, and
independently sampled from a Gaussian distribution
\begin{equation}
J_{klmn}\in \mathcal{N}\left( 0,\frac{3!}{(2N)^{3}}J^{2}\right) ,  \label{J}
\end{equation}%
where $J^{2}=\frac{1}{3!}\sum_{lmn}\left\langle J_{klmn}^{2}\right\rangle $ is sometimes set to $J=1$ for convenience,  c.f.~Ref.~\cite{Cotler2017}.

One can represent $2N$ Majorana Fermions in terms of $N$
Dirac Fermions,  obeying the normal anti-commutation relations $\{c_{j},c_{k}\}=0$, $\{c_{j},c_{k}^{\dag }\}=\delta _{jk}$,
\begin{equation}
\left\{
\begin{array}{l}
c_{j}=\frac{1}{2}(\chi _{2j-1}+i\chi _{2j}) \\
c_{j}^{\dag }=\frac{1}{2}(\chi _{2j-1}+i\chi _{2j})%
\end{array}%
,\right.
\end{equation}%
which can be further expressed by spin-1/2 operators \{$\sigma ^{x},\sigma
^{y},\sigma ^{z},I$\} through  a Jordan-Wigner transformation
\begin{equation}
\left\{
\begin{array}{l}
\chi _{2j-1}=\underset{j-1}{\underbrace{\sigma ^{z}\otimes \cdots \otimes
\sigma ^{z}}}\otimes \sigma ^{x}\otimes \underset{N-j}{\underbrace{I\otimes
\cdots \otimes I}} \\
\chi _{2j}=\underset{j-1}{\underbrace{\sigma ^{z}\otimes \cdots \otimes
\sigma ^{z}}}\otimes \sigma ^{y}\otimes \underset{N-j}{\underbrace{I\otimes
\cdots \otimes I}}%
\end{array}%
.\right.  \label{SYK-spin}
\end{equation}%

In the limit of large number of particles $N \rightarrow \infty$, the Hamiltonian spectral density of $H_0$ has been shown to be a Gaussian, while for finite $N$ the density deviates outside the support of the Gaussian and is well approximated by a Q-Hermite form \cite{Garcia-Garcia2017PRD}.  The ground-state $E_0$, associated with the thermodynamic properties of the system in the low temperature limit \cite{Garcia-Garcia2017PRD,  jevicki2016,  maldacena2016}, is expected to be proportional to $N$,  due to the fermionic nature of the model.  
The spectrum of the ED Liouvillian $ \mathbb{L}_{\text{ED}}^{(2)} $ in \eqref{liouvilleED} only depends on the energy gaps,  laying on the parabola $ \lambda = -\frac{\gamma}{2} E_{nm}^2 + i  E_{nm}  $, with a probability distribution given from the density of gaps of the Hamiltonian $H_0$.  When the jump term is removed,  the eigenvalues $\lambda$ of $\mathbb{L}^{(2)}$ in \eqref{liouvilleBNGLk}  are spread on a two dimensional locus defined by three boundary parabolas parameterized by the ground-state $R= \abs{E_0}$ of $H_0$,  rigidly shifted with time by the trace preservation scalar term in Eq.~\eqref{rhotDBNGL}.

%
	\begin{SCfigure}
 	\centering
	\caption{\label{Fig2DecoTTbar} {\bf Spectral loci of BNGL for the SYK model.}
Eigenvalue density distribution on the complex plane (from white to green) of the BNGL, $\kappa=2$ Liouvillian \eqref{liouvilleBNGLk} and distinct eigenvalues (orange) of the ED Liouvillian \eqref{liouvilleED}.  
In both models the dephasing strength is taken as $\gamma=1$.
The eigenvalues of the undeformed SYK Hamiltonian $H_0$ for $2 N = 26$ Majorana Fermions were calculated with exact diagonalization.
The spectral density of the BNGL, $\kappa=2$ Liouvilian can be approximated by the product of two identical Q-Hermite forms, deformed to parabolas (gray solid lines), centered on the spectral locus of the ED model.  The theoretical boundary (dashed grey line) is given by Eq. \eqref{bound}.  
}
\includegraphics[width=0.43 \linewidth]{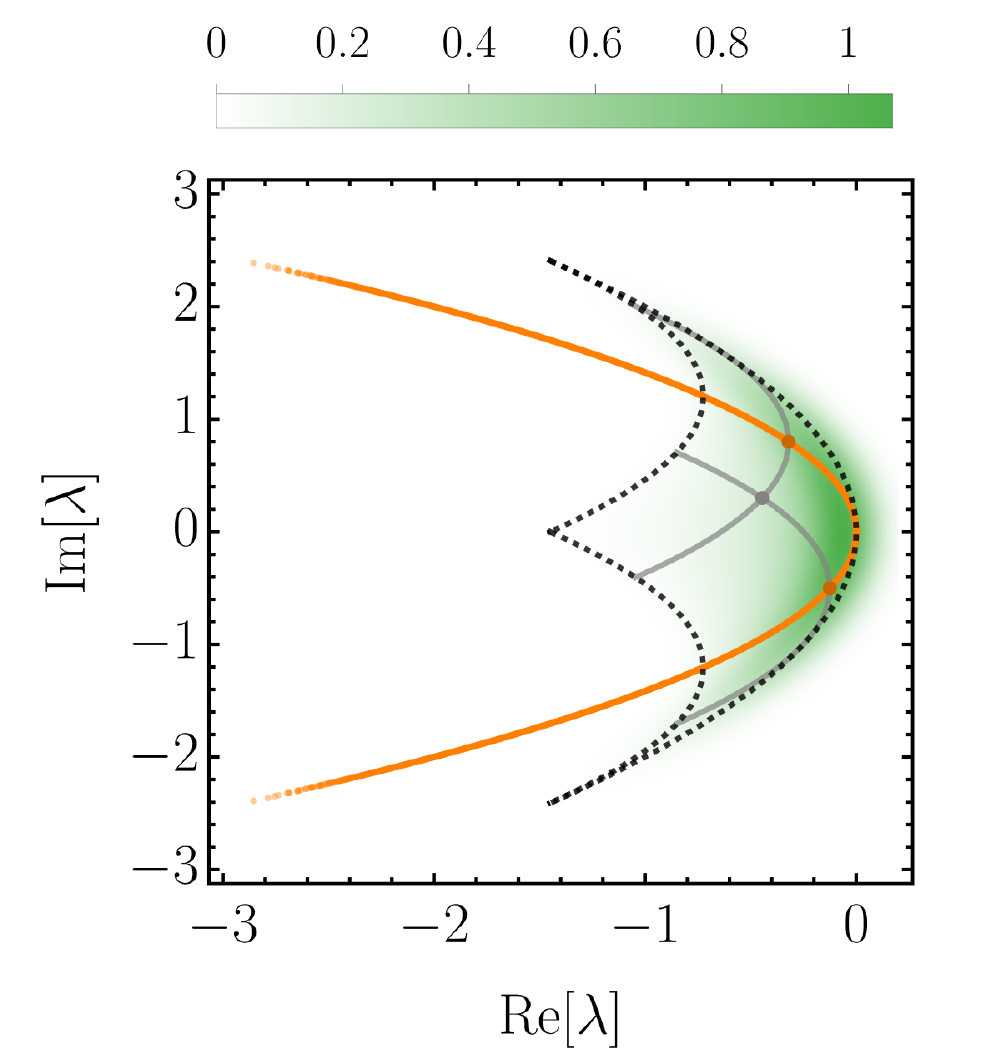}
\hspace{0.5 cm}
\vspace{-0.2 cm}
	\end{SCfigure}

In Fig.~\ref{Fig2DecoTTbar} the spectral density of a BNGL, $\kappa=2$, $\gamma =1$ Liouvillian of $26$ Majorana Fermions is plotted, together with the corresponding spectrum of ED. 
The eigenvalues of the undeformed SYK Hamiltonian $H_0$ were calculated with exact diagonalization.
From Eq. \eqref{liouvilleED} it becomes evident that the spectral density of ED is given by the Hamiltonian spectral density of $H_0$,  deformed to a parabola on the complex plane, $(-\frac{\protect\gamma }{2}E^{2},E)$, while the exclusion of the jump operators spreads the superoperator eigenvalues in a two dimensional locus.  
In general, as shown in  Fig.~\ref{Fig1DecoTTbar} ({\sf{\textbf{d}}},{\sf{\textbf{e}}},{\sf{\textbf{f}}}), for even $\kappa$, given the spectral density of the undeformed Hamiltonian, one can construct the corresponding spectral density of the BNGL by the product of two identical undeformed densities, deformed on their $\kappa$th power, centered on the spectral locus of the ED model.

\subsection{Correlation Dynamics with Thermofield Double Initial  State} \label{SecTFD}

To further illustrate the relevance of non-Hermitian deformations in quantum dynamics we next discuss the evolution of correlations of an entangled state describing two identical copies of a system.
We focus on the thermofield dynamics, initially introduced 
in the study of statistical field theories at finite temperature \cite{takahashi1996}. 
In the search for a formalism where statistical thermal averages can be calculated without trace operations, the TFD of inverse temperature $\beta$ was defined as
\begin{equation}
\label{TFDket}
\ket{\text{TFD}} =  \sum_{n=1}^d \frac{ e^{-\frac{\beta}{2}E_n}}{\sqrt{Z(\beta)}} \ket{n, n}.
\end{equation}
By that time, Bardeen, Carter and Hawking \cite{bardeen1973} had already introduced ``black hole mechanics'', putting forward the consideration of the surface gravity of an axisymmetric  stationary solution of Einstein equations as an analogue of temperature.
Soon after, thermofield dynamics was used by Israel \cite{israel1976} to formalize the ``hot'' thermal vacuum observed outside the horizon of a single radiating eternal black hole.
More recently, thermofield dynamics has been extensively used in the context of AdS/CFT correspondence for the description of contemporaneous black hole pairs in disconnected spaces \cite{maldacena2003,maldacena2013}.
Furthermore, as we will discuss in more detail in Sec. \ref{SecSFF}, the survival probability of a TFD in an isolated system is related to the spectral form factor, a powerful tool in the study of dynamical signatures of quantum chaos \cite{Xu21SFF,cornelius2022}.
In this section we focus on quantum informational quantities, namely the R\'enyi entropy and the logarithmic negativity of a bipartite system, when an initial TFD is evolving under the dynamics generated by non-Hermitian Hamiltonian deformations.

In isolation, the dynamics of two identical non-interacting systems is governed by the Hamiltonian $\tilde H_0 = H_0 \otimes \id  + \id \otimes H_0$.
In the absence of interactions, the entanglement between the two copies is preserved. 
Let us consider the evolution of the whole bipartite system, obeying the dynamics given by the non-Hermitian deformation  $ \tilde H = \tilde H_0 - i \gamma \tilde H_0^2 $. 
The square of the Hamiltonian $\tilde H_0$, describing two identical non-interacting systems is
\begin{align} \label{hamsquare}
\tilde H_0^2 = H_0^2 \otimes \id  + \id \otimes H_0^2 +2 H_0 \otimes H_0 .
\end{align}

By making use of Eq.~\eqref{evolstate} and the  kernel in Eq.~\eqref{dephkernel}, the time-dependent density matrix of an initial TFD (\ref{TFDket}) can be written in the energy eigenbasis of the undeformed Hamiltonian $H_0$ as

    \begin{equation}
    \label{TFDrhot}
 \rho_{\text{TFD}} (t) = \frac{ 
  \displaystyle \sum_{n,m=1}^d  
e^{ -\frac{\beta}{2}(E_n+E_m) - i  2 t ( E_m - E_n)  - 4 \gamma t  \big( E_n^2 +E_m^2 \big )   } 
 }
 { 
\displaystyle \sum_{ \nu=1}^d  
e^{ - \beta  E_\nu - 8 \gamma t E_\nu^2 } 
 }  \dyad{n,n}{m,m} .
    \end{equation}

\subsubsection*{R\'enyi Entropy}
To characterize the evolution of quantum correlations in Eq. (\ref{TFDrhot}), we resort to the R\'enyi entropy.  
For $\alpha \in \mathbb{N}$, the $\alpha$-th power of the reduced density matrix, when the partial trace 
is taken over the second subsystem, is given by
	\begin{align}
\rho^\alpha_{1} (t) = \frac{ 
  \displaystyle \sum_{n=1}^d  
e^{ - \beta \alpha  E_n - 8 \gamma \alpha t E_n^2 } 
 }
 { 
\left( \displaystyle \sum_{ \nu=1}^d  
e^{ - \beta  E_\nu - 8 \gamma t  E_\nu^2 } \right)^\alpha
 } \ketbra{n} .
	\end{align}
For $\alpha \geq 2$, the $\alpha^{th}$  R\'enyi entropy of the subsystem can be written as
     \begin{align} \label{renyiratio}
S_{1,\alpha} ( t ) &=  
\frac{ 1 }{ 1- \alpha } \ln     (
\frac{ 
\displaystyle \sum_{n=1}^d  
e^{ - \beta \alpha  E_n - 8 \gamma \alpha t E_n^2 } 
 }
 { 
\left( \displaystyle \sum_{ \nu=1}^d  
e^{ - \beta E_\nu - 8 \gamma t  E_\nu^2 } \right)^\alpha
 } ) .
     \end{align}
Using the Hubbard–Stratonovich transformation,  the above R\'enyi entropies can be found in terms of the partition function of the undeformed theory
      \begin{align}
 S_{1,\alpha} ( t )  &=   
 \frac{ 1 }{ 1- \alpha } \ln     (
  \frac{1}{\sqrt{\alpha}(32 \pi \gamma  t )^{ \frac{ 1-\alpha }{ 2 } }} 
  \frac{
 \displaystyle  \int_{-\infty}^\infty \mathrm{d}y e^{ -\frac{y^2}{32 \alpha \gamma t} } Z_0(\alpha \beta + i y)
  }
  {
 \left( \displaystyle  \int_{-\infty}^\infty \mathrm{d}y e^{ -\frac{y^2}{32 \gamma t} } Z_0( \beta + i y) \right)^\alpha
  } ) .
      \end{align}
In isolation, the system R\'enyi entropy remains constant and equal to the initial value
\begin{equation}
    S_{1,\alpha} ( 0 ) = \frac{1}{1-\alpha} \ln \left( \frac{Z_0 (\alpha \beta)}{Z_0 (\beta)^\alpha}\right) .
\end{equation}

A remarkable fact is that $S_{1,\alpha}(t) \rightarrow 0$ at large times.
The   R\'enyi-2 entropy ($\alpha=2$) is related to the purity by $P_1(t) = e^{S_{1,2}(t)} \rightarrow 1$.
The first copy of the TFD is becoming asymptotically a pure state with time, and  thus the two copies are  disentangled on this limit and described by a product state.

\begin{figure}[t]
\begin{center}
\hspace{-0.5cm}
\includegraphics[width=0.92 \linewidth]{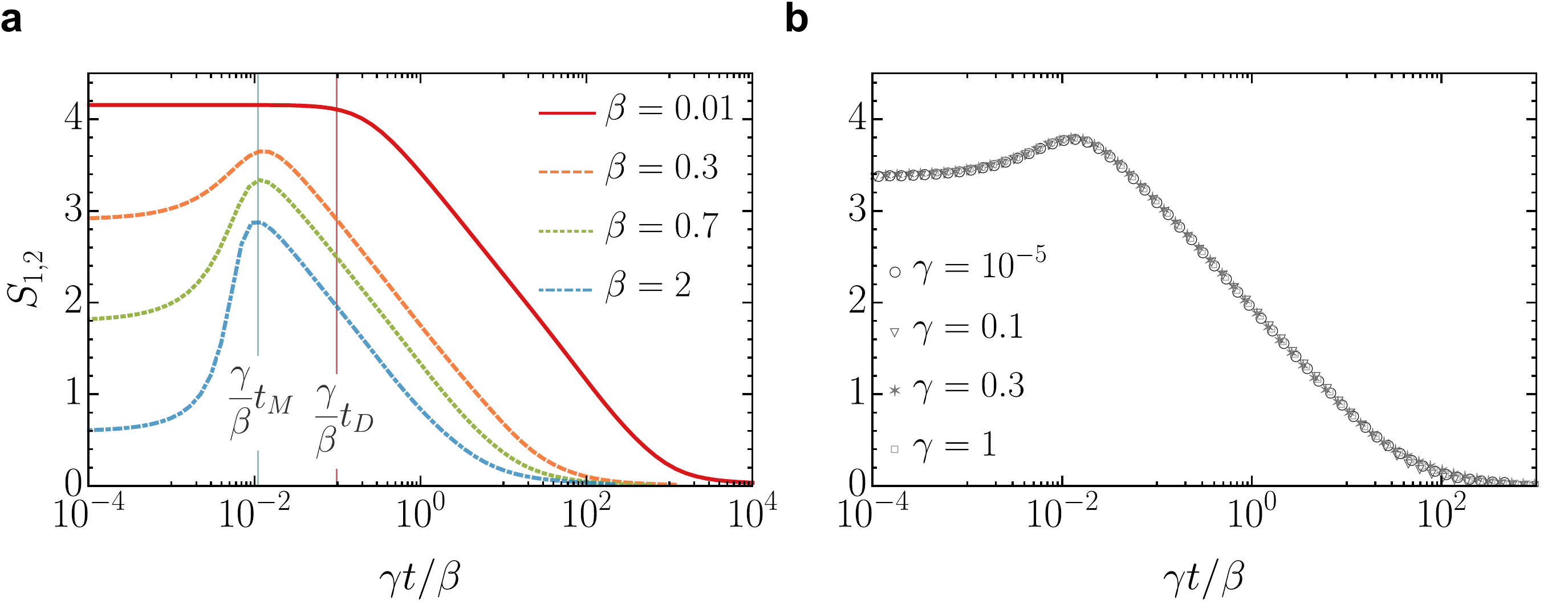}
\end{center}
\vspace{-0.5cm}
\caption{\label{Fig3DecoTTbar} {\bf  R\'enyi-2 entropy of the thermofield double state. }
Hamiltonian averages of $100$ {\rm GOE}(64) Hamiltonians in rescaled time.
{\sf{\textbf{a}}}: At high temperature, $\beta \rightarrow 0$, the R\'enyi entropies start decaying to zero after time $t_D$. 
The two copies of the TFD are effectively being disentangled to a product state.
At low temperatures, the decay comes after the growth to maximum value around a timescale $t_M$, until which entanglement increases.
The two regimes are separated by the critical temperature scale $1/\beta_c$.
{\sf{\textbf{b}}}: R\'enyi-2 entropy for $\beta=0.2$.
In the rescaled time the structure of the R\'enyi entropies rely only on the value of $\beta$.
}
\end{figure}

When the ground state  energy is non-negative, the R\'enyi entropies decrease monotonically.
Contrarily to this, as shown in Fig.~\ref{Fig3DecoTTbar} for samples of {\rm GOE}($d$) Hamiltonians, at finite temperature the R\'enyi entropies can display a single maximum when the energy spectrum contains negative eigenvalues.
For finite low temperature, the R\'enyi entropies grow to the maximum value after which they converge to zero monotonically.
In that case, the short time behavior of the R\'enyi entropy is governed by the exponential of the smallest negative eigenvalue $E_M$.
Specifically, the timescale at which the positive exponents start vanishing in the argument of Eq. \eqref{renyiratio}, given by $\beta E_M + 8 \gamma t_M E_M^2=0$, provides a good approximation for the maximum of the R\'enyi entropy
\begin{equation}
    t_M = \frac{\beta}{8 \gamma E_M} . 
\end{equation}
For a sample of {\rm GOE}($d$) Hamiltonians the average minimum negative eigenvalue can be approximated by the radius of the semicircle law of Eq.~\eqref{semicircle}, $\langle E_M \rangle_H = \sigma \sqrt{2 d}$, and
    \begin{equation}
        t_M = \frac{\beta}{8 \gamma \sigma \sqrt{2 d}}  .
    \end{equation}
The short time behavior in high temperature $\beta \rightarrow 0$ is dictated by the critical timescale $t_D$ at which $8 \gamma  t_D \langle H_0^2 \rangle \simeq 1$ which is connected to the Zeno \cite{delcampo17} and decoherence timescales through $t_D = 8 \tau_D =  \tau_Z^2/(8 \gamma)$.
Specifically, for a sample of {\rm GOE}($d$) Hamiltonians it can be approximated by
    \begin{equation}
        t_D = \frac{1}{16 \gamma \sigma^2 d} .
    \end{equation}
Remarkably, the critical inverse temperature at which $t_M = t_D$ is independent of the dephasing strength $\gamma$ and only relies on the characteristics of the Hamiltonian ensemble, namely
    \begin{equation}
        \beta_c = \frac{1}{\sigma \sqrt{2d}} .
    \end{equation}
In Fig.~\ref{Fig3DecoTTbar} we show the Hamiltonian averages of the second R\'enyi entropy for a sample of $100$ {\rm GOE}(64) Hamiltonians in rescaled time to illustrate the universality of the above result.

%

\subsubsection*{Logarithmic Negativity}

For an alternative characterization of quantum correlations, we resort to the logarithmic negativity $LN[\rho]$, proposed as a non-convex entanglement monotone with an operational interpretation that sets an upper bound to distillable entanglement \cite{vidal2002,plenio05}.
It is defined in terms of the partial transpose $\rho^{\rm PT} $ of the bipartite density matrix as
\beqa LN[\rho]= \log_2 ( \tr\norm{\rho^{\rm PT} }_1) .
\eeqa

The logarithmic negativity of the time-evolution of the TFD, in Eq. (\ref{TFDrhot}) reads
\beqa
\label{LNtTFD}
 LN[\rho_{\text{TFD}}(t)] = \log_2\frac{ 
  \displaystyle 
  \left|\sum_{n=1}^d  
e^{ -\beta E_n/2 - \gamma 4 t E_n^2}\right|^2 
 }
 { 
\displaystyle\sum_{ \nu=1}^d  
e^{ - \beta E_\nu - 8 \gamma t  E_\nu^2 }
}.
\eeqa
The resulting logarithmic negativity carries all the characteristics of the R\'enyi entropies calculated earlier, strengthening our results for the evolution of the entanglement properties of an initial TFD state evolving under BNGL. 
To check that, one can observe that the expression \eqref{renyiratio} for $\alpha \rightarrow 1/2$ differs from \eqref{LNtTFD} only by a multiplicative factor.
We  observe that for the TFD, the logarithmic negativity is related to $S_{1,1/2} (t)$ as $LN[\rho_{\text{TFD}}(t)]=\ln(2)S_{1,1/2}(t)$ and we  defer from a further characterization of it.

Before closing this section, we recall that the above discussion is based on the global deformation of a bipartite system, initially prepared in a TFD.
One could be tempted to assume that the reduction of entanglement is due to the induced interaction term $H_0 \otimes H_0$ of Eq. \eqref{hamsquare}.
Nevertheless, even if deforming only the local Hamiltonian, i.e., $\tilde H = H \otimes \id  + \id \otimes H$, with $H = H_0 - i \gamma H_0^2$,
when the interaction term is absent in the Liouvillian, the R\'enyi entropies and the logarithmic negativity of the TFD state behave similarly. 
Specifically, the corresponding expressions for the local deformation are equal to the ones obtained by the global transformation for half the dephasing strength.

\subsection{Deformation of the Spectral Form Factor}\label{SecSFF}

In the characterization of quantum chaos in terms of the spectral properties of the Hamiltonian describing an isolated quantum system, the correlation between eigenvalues plays a crucial role.  
For any initial pure state undergoing unitary evolution, the Fourier transform of the survival probability (auto-correlation function, two-point correlation function or fidelity between initial and final state) is a weighted sum of $\delta-$functions positioned at the eigenvalues of the Hamiltonian.
Inversely, the absolute square value of the Fourier transform of the local density of states is the survival probability of the initial quantum  state.
When the probability amplitudes of the initial state are the square root of the Boltzmann factors, the survival probability is known as the spectral form factor and provides a convenient tool to characterize dynamical signatures of quantum chaos \cite{Leviandier86,WilkieBrumer91,Alhassid93,Ma95,brezin1997,Gorin06,delcampo17}. 
The partition function with a complex-valued inverse temperature can be considered as a generalization, involving a complex Fourier transform instead \cite{Dyer2017,Cotler2017,delcampo17}. 
The SFF and its generalization exhibit key features as a function of time that include a decay to a minimum value known as the correlation hole, a subsequent growth characterized by a ramp linear in time, and saturation to an asymptotic plateau value.
The depth and area of the correlation hole have been shown to measure the long- and short-range correlations of the energy levels \cite{Ma95}.
This behavior is better appreciated in an ensemble of Hamiltonians, though it can be manifested as well in a single self-averaging system. 
The features of the SFF under Hermitian Hamiltonian deformations have been studied in Ref. \cite{HeLau22}.

In open quantum systems, different quantities have been proposed to characterize the interplay of quantum chaos and decoherence using spectral properties \cite{Haake,Gorin06,Jacquod09,Xu19,Can19,Xu21SFF,li2021,cornelius2022}. 
An analogue of the SFF is given by the fidelity between a coherent Gibbs state 
 \beqa
|\psi _{\beta }\rangle =\sum_{n=1}^d \frac{e^{-\beta E_{n}/2}}{\sqrt{Z_0(\beta )}}|n\rangle,\quad Z_0(\beta)=\tr[e^{-\beta H_0}],
\eeqa
and its time-evolution \cite{Xu21SFF,cornelius2022}. 
Provided that the latter is described by a quantum channel $\Lambda$, the state at time $t$ is given by a density matrix $\rho (t)=\Lambda[ \rho (0) ]$.  
The analogue of the SFF then reads \cite{Xu21SFF,cornelius2022}
\beqa
F (t)=\langle \psi _{\beta }|\rho (t) |\psi
_{\beta }\rangle =\langle \psi _{\beta }\left|\Lambda\left[|\psi_{\beta }\rangle\langle \psi _{\beta }|\right]\right|\psi
_{\beta }\rangle .
\eeqa
In the case of  unitary dynamics  generated by $H_0$,  one recovers the familiar expression
\beqa
F (t)=|Z_0(\beta +it)/Z_0(\beta )|^{2}.
\eeqa

It has been pointed out that decoherence suppresses the dynamical manifestations of quantum chaos in the ED case of Eq. \eqref{rhotED}, i.e., it shrinks the correlation hole of the proposed SFF \cite{Xu21SFF}. 
By contrast, the corresponding dynamics of  the BNGL equation for the deformed Hamiltonian $w(H_0)=H_0-i\gamma H_0^2$ can enhance the aforementioned signatures of quantum chaos \cite{cornelius2022}.
Furthermore, BNGL dynamics leads to an extension of the ramp's span while lowering the values of the dip and plateau,  providing  an experimentally-feasible physical mechanism for the kind of spectral filtering often used in numerical studies of many-body systems \cite{cornelius2022}. 

Let us recall some results from Refs. \cite{Xu21SFF,cornelius2022}. 
The explicit expression of the SFF under Lindbladian ED  (\ref{rhotED}) reads
\beqa
F(t) =\frac{1}{Z_0(\beta)^2}\sum_{n,m=1}^{d} e^{-\beta(E_n+E_m)-it(E_m-E_n)-\gamma t (E_m-E_n)^2}.
\eeqa
Note that this equation also describes the fidelity for an initial TFD, by time rescaling \cite{Xu19,delCampo2020,Xu21SFF}.
Importantly, it can be written in terms of the partition function analytically continued to complex inverse temperature as  \cite{Xu19,delCampo2020,Xu21SFF}
\beqa
F(t)=\sqrt{\frac{1}{4\pi \gamma t}} \int_{-\infty }^{\infty }\mathrm{d}ye^{-\frac{y^{2}}{4\gamma t}}\left|\frac{Z_0 \big(\beta+i(y+t) \big)}{Z_0(\beta)}\right|^2,
\eeqa
thus facilitating its study in cases  in which  the partition function is readily available, e.g.,  in certain integrable models and conformal field theories.
For $t\gg \tau_D$,  the time-evolving density matrix is effectively diagonal and
\beqa
\label{longtF}
F (t) \sim F_p=\frac{1}{Z_0(\beta)^2}\sum_nN_ne^{-2\beta E_n}\geq
 \frac{Z_0(2\beta)}{Z_0(\beta)^2},
\eeqa
where $N_n$ is the degeneracy of the eigenvalue $E_n$.

By contrast, under the BNGL evolution 
a direct application of Eq.~\eqref{evolstate} yields
\begin{equation}\label{fidelityFede}
    F(t)
    =
    \frac{\left| \int_\mathbb{R}  \text{d}s \,    K_w(t,s) 
		Z_0(\beta +i s)\right|^2}{ Z_0(\beta ) \int_\mathbb{R}  \text{d}s \, \int_\mathbb{R}  \text{d}s'\,   K_w(t,s) K_{w^*}(-t,-s')
		 Z_0(\beta +i (s-s'))}\,.
\end{equation}
In the special case of $w(z)=z-i\gamma z^2$ it reads \cite{cornelius2022}
\beqa
\label{FtBNGL}
F(t)=\frac{\left|\displaystyle \sum_{n=1}^d  e^{-(\beta+it)E_n-\gamma t E_n^2}\right|^2}{Z_0(\beta) \displaystyle \sum_{j=1}^d e^{-\beta E_j-2t\gamma E_j^2}}
\,.
\eeqa
\begin{figure}[t]
\begin{center}
\hspace{-0.4cm}
\includegraphics[width=1 \linewidth]{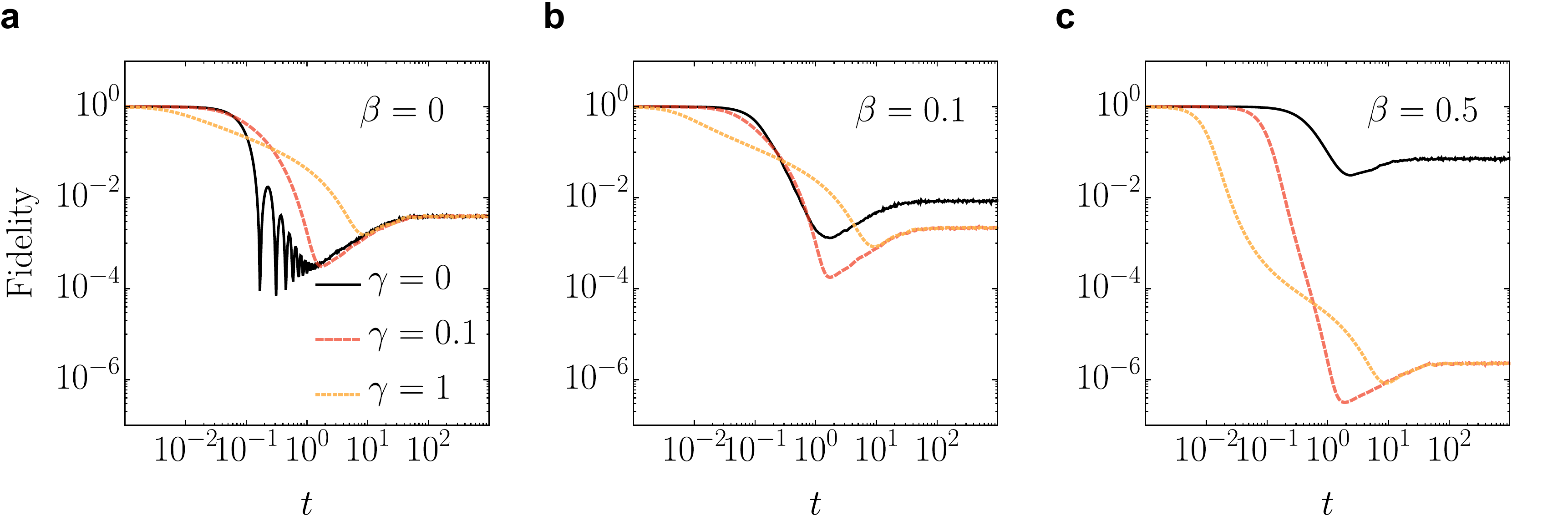}
\end{center}
\vspace{-0.5cm}
\caption{\label{Fig4DecoTTbar} {\bf  The deformed SFF for BNGL in random matrix theory.} The time-dependence of the fidelity between a coherent Gibbs state and its time evolution generalizes the notion of the SFF of Hermitian Hamiltonians to open quantum systems, including those governed by non-Hermitian Hamiltonians.
Different panels correspond to different temperatures of the initial state.  Each panel shows the evolution under BNGL for $\kappa=2$ and different values of the dephasing strength.
The average is taken over a sample of $1000$  ${\rm GOE}(2^8)$ undeformed Hamiltonians $H_0$.
At infinite and high temperature, the nonunitary dynamics under BNGL preserves the main features of the SFF, displaying a decay,  a dip,  a ramp and a plateau.
Deviations from the unitary case ($\gamma=0$), suppress quantum noise in the neighborhood of the dip as well as in the plateau. Increasing the dephasing strength $\gamma$ alters the decay, delaying the appearance of the dip and shortening the ramp, while keeping the onset of the plateau unaltered. 
At lower temperatures (when the annealed approximation is expected to fail), BNGL prolongs the decay, enhancing the dip.  As a result the ramp and the plateau take lower values than in the unitary case.
}
\end{figure}

In Fig.~\ref{Fig4DecoTTbar} we show examples of the characteristic behavior of the deformed SFF \eqref{FtBNGL}, for averages over different Hamiltonian matrices drawn from the Gaussian orthogonal ensemble when the Hilbert space dimension is $d=256$.

The long-time limit of the fidelity, for any $\gamma>0$, reads $F(t)\geq 1/Z_0(\beta)$ where the inequality is saturated for systems lacking degeneracies, e.g., exhibiting quantum chaos \cite{cornelius2022}.
By contrast, for $\gamma=0$,  the value of $F_p$ under ED is given by Eq. (\ref{longtF}).

The choice of the coherent Gibbs state $|\psi_\beta\rangle$ also allows to illustrate, somewhat dramatically, the different nature of the dissipative dynamics in  the presence and absence of the quantum jump term.
To this end, consider the evolution of the purity for an initial coherent Gibbs state.
Under ED  \cite{Xu19,delCampo2020},
\beqa
P(t)=
\sum_{nm}\frac{e^{-\beta(E_n+E_m)}}{Z_0(\beta)^2}e^{-2\gamma t (E_n-E_m)^2}\nonumber=\sqrt{\frac{1}{8\pi \gamma t}}%
\int_{-\infty }^{\infty }\mathrm{d}ye^{-\frac{y^{2}}{8\gamma t}}\left|\frac{Z_0(\beta+iy)}{Z_0(\beta)}\right|^2.
\eeqa
By contrast, as previously mentioned, for an initial coherent Gibbs state evolving under BNGL, the purity remains equal to unity at all times $P(t)=1$.
We also note that if the initial state is mixed, in both cases the purity varies as a function of time, according to (\ref{PtED}) and (\ref{PtEDBNGL}).

In short, for a coherent Gibbs state in the cases of ED with a single Lindblad operator and the corresponding evolution with BNGL, we have been able to express the fidelity and the purity in terms of the partition function of the undeformed Hermitian Hamiltonian using non-Hermitian deformations. 

\begin{figure}[t]
\begin{center}
\hspace{0.0cm}
\includegraphics[width=1 \linewidth]{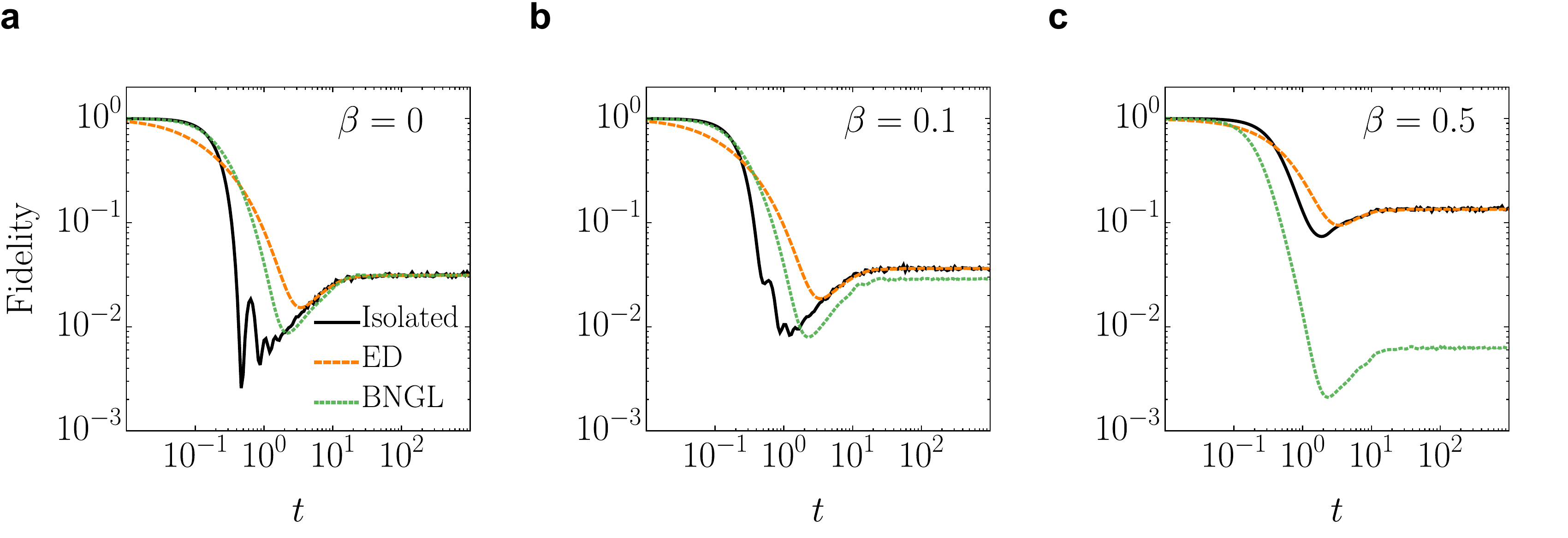}
\end{center}
\vspace{-0.5cm}
\caption{\label{Fig5DecoTTbar} {\bf  Energy-Dephasing vs BNGL  in Random Matrix Theory. } The time-dependence of the fidelity between a coherent Gibbs state and its time evolution is compared for energy-dephasing (orange dashed line) and the  nonlinear evolution for null-measurement conditioning associated with BNGL (green dotted line) with $\gamma=0.2$. For the sake of comparison, we also show the SFF of the corresponding isolated system (black solid line), $\gamma=0$. We consider a sample of 1000 independent Hamiltonians $H_0$ taken from ${\rm GOE}(2^5)$.  At long times,  the fidelity reaches a plateau with the value $\la Z(2\beta)/Z^2(\beta)\ra$ for $\gamma=0$ and $\la 1/Z(\beta)\ra$ for $\gamma\neq 0$.}
\end{figure}

To explore the extent to which the SFF for Eq.~(\ref{rhotED}) and BNGL equation differ, let us assume that $H_0$ is a chaotic Hermitian Hamiltonian with time-reversal symmetry sampled from $H_0\in {\rm GOE}(d)$. 
Specifically, in Fig.~\ref{Fig5DecoTTbar}, we sample $1000$ Hamiltonians with $d=32$, choosing dephasing strength $\gamma = 0.2$ for different temperatures.

%

%


For the non-Hermitian Hamiltonian deformations defined in Eq.~\eqref{deformk} the deformed SFF becomes
\begin{align} \label{SFFgen}
F(t) &= \frac{\displaystyle \sum_{n=1}^d p_n^2 e^{-2 E_n^\kappa t} }{\displaystyle \sum_{j=1}^d p_j e^{-2 E_j^\kappa t}} 
+ 2 \frac{ \displaystyle \sum_{\substack{n,m=1 \\ n<m}}^d p_np_m e^{-(E_n^\kappa+E_m^\kappa) t}   
\cos \Big( \big( E_m -E_n \big) t \Big) }
{\displaystyle \sum_{j=1}^d p_j e^{-2 E_j^\kappa t}},
\end{align}
where $p_n=e^{-\beta E_n}/Z_0(\beta)$ are the Boltzmann factors of the undeformed Hamiltonian $H_0$.

The timescale at which all frequencies $E_m-E_n$ have on average been expressed in the evolution can be approximated by the inverse of the average level spacing $\Delta$, sometimes referred to as Heisenberg time $t_H = 2 \pi/ \Delta$ \cite{prange1997,Haake}.
After this time, the cosines of frequencies whose ratio is irrational cancel each other on average, leading the SFF to its plateau, while the distribution of the smallest ones, i.e., the level spacing distribution determines its behavior right before $t_H$.
In a quantum chaotic system, level repulsion is manifested in the ramp which follows the dip of the correlation hole leading the SFF to saturation.
In this context the absence of a correlation hole before the Heisenberg time is associated with regular dynamics.
The mean level spacing for a Hamiltonian sampled from $\mathrm{GOE}(d)$, whose spectrum has not been unfolded, is $\Delta  = \sigma \sqrt{8 d} / (d-1)$, and thus the Heisenberg time, $t_H = \pi  (d-1) / ( \sigma \sqrt{2d} )$.
%
\begin{figure}[t]
\begin{center}
\hspace{-0.5cm}
\includegraphics[width=1 \linewidth]{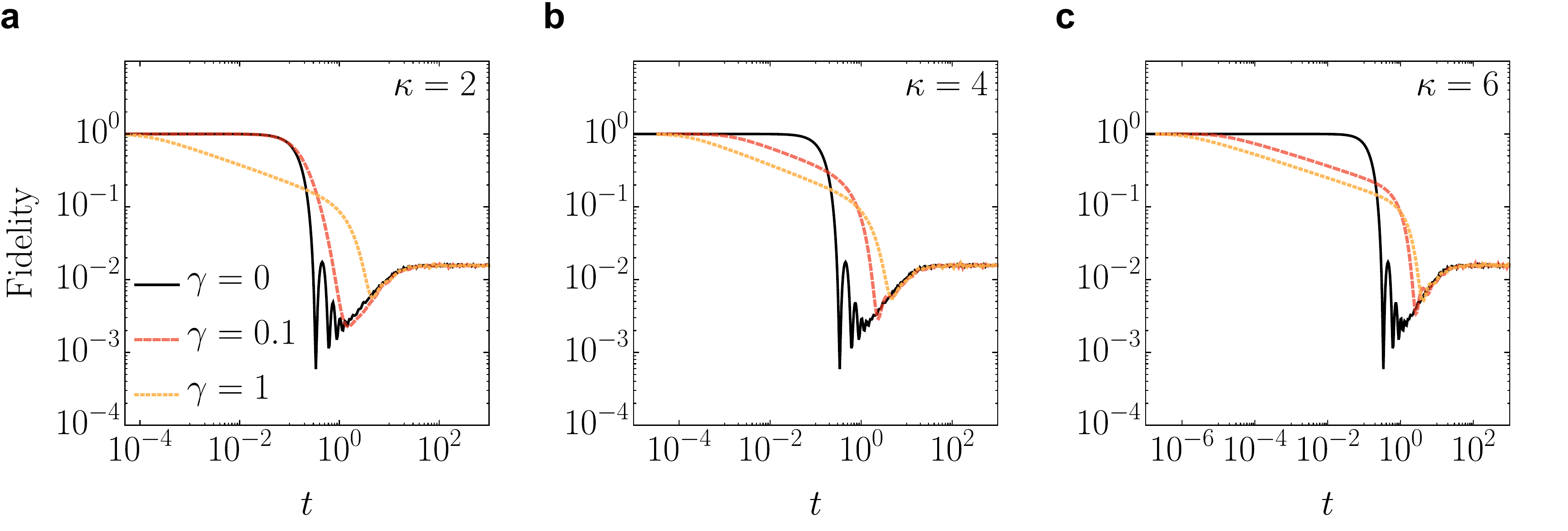}
\end{center}
\vspace{-0.5cm}
\caption{\label{Fig6DecoTTbar} {\bf  Correlation hole shrinking for different non-Hermitian deformations. }
Fidelity between initial and time evolved infinite inverse temperature coherent Gibbs state under the BNGL dynamics of $\mathbb{L}^{(\kappa )}$, for even $\kappa$ and characteristic values of the dephasing strength $\gamma$.  Three different deformations for $\kappa =2,4,6$ are averaged over a sample of $1000$  ${\rm GOE}(2^6)$ undeformed Hamiltonians $H_0$ with $\sigma =1$. 
}
\end{figure}

In the non-Hermitian Hamiltonian deformations of Eq.~\eqref{deformk}, the dissipative part of the Liouvillian commutes with the system Hamiltonian, leaving the frequencies in Eq.~\eqref{SFFgen} unaffected.
Namely, the $H_0^\kappa$ part of the deformation affects the depth and area of the correlation hole.
In Fig.~\ref{Fig6DecoTTbar} we show the shrinking of the correlation hole with the increase of the dephasing strength, while the Heisenberg time remains unchanged. 
In all three panels we show the Hamiltonian averages of $1000$ Hamiltonians, sampled from ${\rm GOE}(64)$ for infinite temperature $\beta=0$ and different values of the dephasing strength.

\section{Liouvillian Deformations}\label{SecLD}

Before closing, we  discuss the generalization of our results to the case of arbitrary open quantum dynamics. The evolution of the quantum state is generated by a Liouvillian $\mathbb{L}_0$, which may be diagonalizable or not.   For simplicity, we focus on the former case. 
Let $\mathbb{L}_0 = \sum_n \lambda_n |n)(\tilde n|$ be a Liouvillian without any exceptional points \cite{MingantiPRA2019,RoccatiOSID2022}, diagonal in a bi-orthogonal basis, after the vectorization process presented in Sec.  \ref{SecSS}, with $|n)$ and $(\tilde n|$ being the right and left eigenstates, respectively, of the complex eigenvalue $\lambda_n$ \cite{BrodyJPA2013,Gyamfi20}.
The equation
\begin{align} \label{vecdyn}
\partial_t |\rho(t)) = \mathbb{L}_0 |\rho(t)) ,
\end{align}
is solved by $|\rho(t)) = \mathbb{\Phi}_0 |\rho (0)) $, for the dynamical map $\mathbb{\Phi}_0 = e^{\mathbb{L}_0 t} = \sum_n e^{\lambda_n t} |n)(\tilde n|$. 
Extending the discussion of Sec.  \ref{SecNHHD} to Liouville space,  a deformation $w$ gives $w(\mathbb{L}_0)=\sum_n w(\lambda_n) |n)(\tilde n|$ and $\mathbb{\Phi}_w = \sum_n e^{w(\lambda_n)t} |n)(\tilde n|$. 
As an example, consider the case in which the undeformed Liouvillian $\mathbb{L}_0$ describes a system in isolation. The Liouvillian  takes the form \eqref{isoliouv} and is thus anti-Hermitian.
The right and left eigenvectors coincide and the eigenvalues are purely imaginary,  given by the frequencies $\omega_n$, which determine the dynamics,  $\lambda_n = - i \omega_n$.
The deformed and undeformed propagators can be obtained through a Fourier transform 
\begin{equation}
    \mathbb{\Phi}_w(t)
    =
    \int_\mathbb{R}  \text{d}t'\,  K_w(t,t')  \mathbb{\Phi}_0 (t'),
\end{equation}
with 
\begin{equation}
	K_w(t,t')
	=
	\int_\mathbb{R}  \frac{ \text{d} \omega }{2\pi}\, e^{i t'  \omega-iw (\omega)t} .
\end{equation}
A simple deformation that preserves the Lindblad structure is 
$w(\mathbb{L_0}) = \mathbb{L_0} + \gamma \mathbb{L_0}^2$, $\gamma \in \mathbb{R}$, starting from a Liouvillian describing an isolated system $\mathbb{L_0} =-i\left(H \otimes \id-\id\otimes H^\intercal \right)$.
In this case, one recovers the full energy dephasing dynamics of Eq. \eqref{rhotED}.
In addition, one can consider the deformation $w(\mathbb{L_0}) = \mathbb{L_0} + \gamma \mathbb{L_0}^{2s}$ with  $s\in\mathbb{N}$, which leads to the master equation
\beqa
\partial_t\rho=-i[H_0,\rho]+(-1)^s\gamma[H_0,[H_0,\cdots,[H_0,\rho]]],
\eeqa
involving a dissipator with $2s$ nested commutators.
While the latter is not manifestly of Lindblad form, it is solved by the quantum state
\beqa 
\rho(t)=\sum_{nm}\rho_{nm}(0)e^{-i(E_n-E_m)t+(-1)^s\gamma t (E_n-E_m)^{2s}}|n\ra \la m|.
\eeqa
Therefore, for odd $s$, the dynamics generalizes  the usual case of energy dephasing ($s=1$). For even $s$, with a bounded spectrum, it has the opposite effect as the time evolution enhances the coherences in the energy eigenbasis. 

This example illustrates the versatility of leveraging the notion of non-Hermitian Hamiltonian deformations to  more general open dynamics. 
Quantum channel deformations at other levels are left for future investigations.

\section{Conclusions}\label{secConclusions}

Integrable and exactly-solvable Hamiltonian deformations constitute a powerful tool among non-perturbative methods. Using them, equilibrium correlations of the deformed theory can be found via integral transforms  in terms of those in the original theory \cite{Gross20,Gross20b}.

In this work, using the theory of open quantum systems, we have motivated the introduction of non-Hermitian Hamiltonian deformations.  In the context of continuous quantum measurements, the latter describe the dynamics of a subensemble of trajectories selected according to a measurement record, i.e., the absence of quantum jumps.
For such subensemble, the dynamics is governed by a non-linear non-Hermitian evolution characterized by balanced norm gain and loss. The spectrum of the deformed non-Hermitian Hamiltonian is no longer real, but complex-valued. This makes it possible to  express nonequilibrium correlations of the deformed theory in terms of those of the undeformed theory, using integral equations, i.e., generalizing the relations known in the Hermitian setting.
In doing so, we have elucidated the relation between the time-evolution operators, density matrices, and the spectral properties of the generators of time evolution in both the deformed and undeformed theories.

We have explored 
the energy dephasing channel under both Markovian and BNGL evolution.
We found that the spectral properties are significantly altered, as the Liouvillian spectrum constitutes a one dimensional locus in the complex plane, while the BNGL spectrum corresponds to a two dimensional one. As an  example, we considered the SYK model and random Hamiltonians from the GOE.
We characterized quantum correlations starting from a thermofield double state. Remarkably, entanglement and R\'enyi entropy display a maximum value under BNGL evolution, scaling with inverse temperature.

As an application of non-Hermitian deformations and building on earlier results, we considered 
signatures of quantum chaos, using the survival probability of a coherent Gibbs state, identifying the effect of quantum jumps. 
The spectral form factor exhibits a decay, dip and plateau. The dip is generally suppressed under energy dephasing, in the presence of quantum jumps. 
However, when conditioning the dynamics to the absence of the latter, we have shown that non-Hermitian evolution of the energy dephasing channel under BNGL can actually enhance signatures of chaos by broadening the duration of the ramp. 
This is contrary to the expectation that decoherence generally suppresses signatures of quantum chaos. 
Further, the value of the plateau is fundamentally distinct from the isolated case and is characterized by the inverse of the partition function in quantum chaotic systems.

Finally, we discuss a possible way to generalize the presented theory of non-Hermitian deformations to Liouvillians, through the introduction of integral kernels which associate the deformed and undeformed dynamical maps.

Beyond these findings, non-Hermitian and Liouvillian deformations should find broad applications in the study of dissipative quantum many-body systems, the interplay between information scrambling and information loss, black hole physics and unitarity breaking,  and gauge-gravity dualities in open systems.

\section{Acknowledgments}
It a pleasure to acknowledge useful discussions with Niklas H{\"o}rnedal, Federico Balducci, Nicoletta Carabba, Pablo Martínez-Azcona and Shinsei Ryu. 

\appendix

\section{Time evolution under BNGL }\label{appFid}

From Eq.~\eqref{kernel}, we have that 
\begin{align}
		K_w(t,s)^*
		&=
		\int_\mathbb{R}  \frac{ \T{d}E}{2\pi} e^{-isE+iw^* (E)t}
		\\
		&=
		\int_\mathbb{R}  \frac{ \T{d}E}{2\pi} e^{i(-s)E-iw^* (E)(-t)}
		\\
		&=
		K_{w^*}(-t,-s)
\end{align}
so that 

\begin{equation}
	U_w^\dagger (t)
	=
	\int_\mathbb{R}  \T{d}s\,  K_{w^*}(-t,-s)  U_0^\dagger (s)
\end{equation}

Therefore

\begin{align}
	\tilde\rho_w(t)
	&=
	U_w (t)\rho(0)  U_w^\dagger (t)
	\\
	&=
	e^{-iw (H_0)t} \, \rho(0)  \,   e^{iw ( H_0)^\dagger t}
	\\
	&=
	\int_\mathbb{R}  \T{d}s \, \int_\mathbb{R}  \T{d}s'\,   K_w(t,s) K_{w^*}(-t,-s')  U_0 (s) \rho(0)  U_0^\dagger (s')
	\\
	&=
	\int_\mathbb{R}  \T{d}s \, \int_\mathbb{R}  \T{d}s'\,   K_w(t,s) K_{w^*}(-t,-s')  U_0 (s) \hat  U_0^\dagger (s')  U_0 (s') \rho(0)  U_0^\dagger (s')
    \\
	&=
	\int_\mathbb{R}  \T{d}s \, \int_\mathbb{R}  \T{d}s'\,   K_w(t,s) K_{w^*}(-t,-s')  U_0 (s-s')\tilde\rho_0(s').
\end{align}

\bibliographystyle{apsrev4-1}  
\bibliography{DecoCFT_lib}

\end{document}